\def\year{2020}\relax
\documentclass[letterpaper]{article} 
\usepackage{aaai20}  
\usepackage{times}  
\usepackage{helvet} 
\usepackage{courier}  
\usepackage[hyphens]{url}  
\usepackage{graphicx} 
\usepackage{graphicx}  
\usepackage{algorithm}
\usepackage[noend]{algpseudocode}
\usepackage{amsmath}
\usepackage{amssymb}
\usepackage{amsthm}
\usepackage{epstopdf}
\usepackage{enumitem}
\usepackage{latexsym}
\usepackage{multirow}
\usepackage{mathtools}
\usepackage{soul}

\newtheorem{thm}{Theorem}
\newtheorem{cor}{Corollary}
\newtheorem{lem}{Lemma}
\newtheorem{prop}{Proposition}
\newtheorem{defn}{Definition}

\newcommand{\M}{\mathcal{M}}

\newcommand{\T}{\mathcal{T}}

\newcommand{\ba}{\mathbf{a}}

\newcommand{\ipomdp}{\textsf{I-POMDP}}
\newcommand{\ipomdplite}{\textsf{IPOMDP-Lite}}
\newcommand{\nestedmdp}{\textsf{Nested-MDP}}

\title{Scalable Decision-Theoretic Planning in Open and Typed Multiagent Systems}
\author{
Adam Eck,\textsuperscript{\rm 1}
Maulik Shah,\textsuperscript{\rm 2}
Prashant Doshi,\textsuperscript{\rm 2},
Leen-Kiat Soh\textsuperscript{\rm 3}\\
\textsuperscript{\rm 1}Oberlin College,
\textsuperscript{\rm 2}University of Georgia,
\textsuperscript{\rm 3}University of Nebraska\\ 
aeck@oberlin.edu, mns28652@uga.edu, pdoshi@uga.edu, lksoh@cse.unl.edu
}
 
\begin{document}

\maketitle

\begin{abstract}
In open agent systems, the set of agents that are cooperating or competing changes over time and in ways that are nontrivial to predict. For example, if collaborative robots were tasked with fighting wildfires, they may run out of suppressants and be temporarily unavailable to assist their peers. We consider the problem of planning in these contexts with the additional challenges that the agents are unable to communicate with each other and that there are many of them. Because an agent's optimal action depends on the actions of others, each agent must not only predict the actions of its peers, but, before that, reason whether they are even present to perform an action.  Addressing openness thus requires agents to model each other’s presence, which becomes computationally intractable with high numbers of agents.  We present a novel, principled, and scalable method in this context that enables an agent to reason about others' presence in its shared environment and their actions. Our method extrapolates models of a few peers to the overall behavior of the many-agent system, and combines it with a generalization of Monte Carlo tree search to perform individual agent reasoning in many-agent open environments.  Theoretical analyses establish the number of agents to model in order to achieve acceptable worst case bounds on extrapolation error, as well as regret bounds on the agent's utility from modeling only some neighbors.  Simulations of multiagent wildfire suppression problems demonstrate our approach's efficacy compared with alternative baselines.
\end{abstract}

\section{Introduction}

In both cooperative and competitive multiagent systems (MAS), a participating agent benefits from reasoning about how other agents will behave while choosing optimal actions that maximize its chances of accomplishing shared or self-interested goals.  However, nuances in real-world environments often challenge straightforward peer modeling. One of these is \textbf{agent openness} occurring whenever individual agents join or leave the system (temporarily or permanently) over time.  For example, cooperative robots tasked with suppressing wildfires alongside or in place of human fire fighters would need to periodically leave the environment to recharge their limited suppressants that were spent during firefighting.  Likewise, competitive autonomous ride-sharing cars can no longer compete for new passengers while transporting a full ride.   Consequently, openness requires that an agent not only predict \emph{what} actions their neighbors will take, but also \emph{whether} they are even present to take actions.  Inaccurately predicting the presence or absence of others may cause agents to choose actions that miss their intended benefits or other utility maximizing opportunities.

Furthermore, real-world environments such as wildfire suppression often involve {\em many agents of various types} working together to put out the wildfires.  Thus, the modeling and reasoning must additionally scale with the number of agents.  As scalability is a general and ever-present challenge for multiagent planning, investigations have first focused on multiagent planning for smaller groups of agents, e.g., \cite{Amato:Scalable,Claes:Warehouse,Hoang13:Interactive,Nair05:Networked,Oliehoek:ExpLocality}. Subsequently, there has been some focus on scalable planning in large groups of agents \cite{Oliehoek:APPXSOL,Nguyen:CDecPOMDP,Sonu:2017,Velagapudi:DistModel} albeit, in the absence of openness.   Indeed, state-of-the-art multiagent planning in open environments has demonstrated successful planning with up to 5 agents~\cite{Chandrasekaran:Open,Cohen:OpenDecPOMDPs}.  Of course, openness compounds the challenges of scalable reasoning since not only should the agent model how its many peers decide their actions, but also how they dynamically join and leave the environment.

In this paper, we focus on achieving scalable multiagent reasoning in open environments.  We improve on a previous approach to individual agent planning in the context of agent openness~\cite{Chandrasekaran:Open} to consider many-agent settings.  To promote flexible scalability, our method enables a subject agent to intelligently select a small subset of neighbors to explicitly model. Then, it extrapolates their expected behaviors to the larger set of all agents. This approach has its roots in surveying and polling theory where surveyors rarely poll all individuals in their target class.  As such, this approach represents a novel integration of survey theory in multiagent planning. However, selective modeling may not be enough for many-agent systems.  We observe that in many open domains such as wildfire suppression, which particular firefighter is performing an action is not relevant to the decision making. Similar observations have been made in other domains ~\cite{Velagapudi:DistModel,Nguyen:CDecPOMDP,Sonu:2017}. Consequently, further efficiency is made possible by exploiting the property of anonymity of the other agents~\cite{Sonu:2017}. These modeling approaches are utilized in a new generalization of Monte Carlo tree search~\cite{Silver10:POMCP} to plan in open, many-agent settings.   

Our results in this paper are both theoretical and experimental. We analytically establish ($i$) the number of neighbors that a subject agent must model to achieve given worst case extrapolation errors, and ($ii$) the regret bounds on the agent's approximate utility function from modeling some neighbors only. To empirically demonstrate scalability and evaluate the benefits of our approach, we extend benchmark simulations of wildfire suppression \cite{Chandrasekaran:Open} to include setups with up to 50 agents -- an order of magnitude larger than the prior literature -- capable of performing up to $1.27 \times 10^{30}$ joint actions, for which existing decision-theoretic planning does not scale.

\section{Background}

We review the planning framework and a previous approach for integrating anonymity into reasoning about others.

\subsection{I-POMDP-Lite Framework}

A framework for individual agent planning in partially observable MAS is the I-POMDP-Lite~\cite{Hoang13:Interactive}. This framework retains many of the benefits of the more general I-POMDP framework but mitigates computational costs by modeling the other agents' reasoning processes as an approximation of their true decision making. Formally,
\begin{align*}
\ipomdplite{}_{i,l}^{\mathcal{L}} \triangleq \langle Ag, S, A, \Omega_i, T_i, O_i, R_i, \gamma, b_{i, 0}, \\
\{\M_{j,l-1}, \M_{k,l-1}, \ldots, \M_{z,l-1}\} \rangle
\end{align*}

\begin{itemize}[leftmargin=*,topsep=0pt,itemsep=0pt]
\item $Ag$ is a set of agents, consisting of a \emph{subject agent} $i$ using the I-POMDP-Lite to decide how to act and other agents $j, \ldots, z$ modeled by subject $i$.  Each agent has a frame $\theta$ from the discrete set $\Theta = \{\theta_1, \ldots, \theta_{|\Theta|}\}$. A frame represents the agent's capabilities and preferences.

\item  $S$ is the set of states of the decision-making problem, possibly factored into  variables $F_1 \times F_2 \times \ldots \times F_k$, such as the intensities of the $k$ wildfires the agents need to suppress. 

\item $A = A_i \times A_j \times \ldots \times A_z$ is the set of possible joint actions of the agents; e.g., the individual fires that each agent chooses to fight.  For notational convenience, $\mathbf{a_{-i}} \in A_j \times \ldots \times A_z$ denotes the vector of actions by $Ag \setminus \{i\}$.

\item $\Omega_i$ is the set of observations of agent $i$.

\item $T_i(s, a_i, \mathbf{a_{-i}}, s') = P(s' | s, a_i, \mathbf{a_{-i}})$ gives the probabilities of stochastic state transitions caused by actions. 

\item $O_i(s', a_i, \mathbf{a_{-i}}, o_i) = P(o_i | a_i, \mathbf{a_{-i}}, s')$ models the probabilities of stochastic observations revealed to subject agent $i$ after joint action $(a_i, \mathbf{a_{-i}})$.

\item $R_i(s, a_i, \mathbf{a_{-i}}) \in \mathbb{R}$  is the reward function of agent $i$ dependent on the state and joint actions.

\item $\gamma \in (0, 1]$ and $b_{i,0}$ are the discount factor and initial belief state of $i$, respectively. 

\item $\{\M_{j,l-1}, \M_{k,l-1}, \ldots, \M_{z,l-1}\}$ is the set of  mental models ascribed to the other agents $j$, $k$,  $\ldots$, $z$ at level $l-1$.~\footnote{The level of a reasoning process represents the level in a hierarchy of agents reasoning about their neighbors: level 1 implies all others choose actions randomly, and level $l > 1$ implies that all others reason at level $l - 1$.}  Each mental model $\M_{j, l-1}$ is a Nested-MDP. Hoang and Low (\citeyear{Hoang13:Interactive}) define a Nested-MDP as:
$$ \M{}_{i,l} \triangleq \langle S, A, T_i, R_i, \{\pi_{j,d}, \pi_{k,d}, \ldots, \pi_{z,d}\}_{d=0}^{l-1}, \gamma \rangle $$
where $\{\pi_{j,d}, \pi_{k,d}, \ldots, \pi_{z,d}\}_{d=0}^{l-1}$ is the
  set of policies followed by other agents $j$, $k$,  $\ldots$, $z$ at levels 0 to $l-1$. The policies are obtained by solving their own Nested-MDPs.
\end{itemize}

\noindent In solving an I-POMDP-Lite, an agent chooses actions that maximize the cumulative, discounted reward function over a finite horizon $H$, $r_0 + \gamma r_1 + \gamma^2 r_2 + \ldots + \gamma^{H-1} r_{H-1}$, by considering Bellman equations for each belief/action pair:

\begin{small}
\begin{align}
&Q_{i,l}^t(b_i, a_i) =   \rho_i(b_i,a_i) + \gamma \sum_{s',o_i} {\T{}_i^{a_i,o_i}(s',o_i|b_i,a_i) V_{i,l}^{t-1}(b_i')}
\label{eqn:Q}
\end{align}
\end{small}
\begin{small}
\begin{align}
V_{i,l}^{t}(b_i') = \max_{a_i \in A_i} Q_{i,l}^t(b_i, a_i)
\label{eqn:V}
\end{align}
\end{small}
where
\begin{align}
\begin{split}
\rho_i(b_i,a_i) = \sum\limits_{s \in S} \sum_{\ba_{-i} \in A_{-i}} & \prod\limits_{-i \in \{j,k,\ldots, z\}} \pi_{-i,l-1}(s,a_{-i}) \\&\times R_i(s,a_i,\ba_{-i})~b_i(s) 
\end{split}
\label{eqn:rho}
\end{align}
Policies $\pi_{-i,l-1}(s,a_{-i})$ are solutions  of the  other
agents' mental models $\M_{-i,l-1}$, and $b_i'$ denotes the updated belief $Pr(s'|o_i,a_i,\ba_{-i},b_i)        \propto O_i(s',a_i,o_i) \sum\limits_{s \in S} T_i(s,a_i,\ba_{-i},s')$ $b_i(s)$.  Hoang and Low (\citeyear{Hoang13:Interactive}) establish regret bounds on the value function $V$ based on any error introduced through approximating other agents as reasoning using a Nested-MDP (instead of their true reasoning process).

Both the I-POMDP-Lite and decentralized POMDP (Dec-POMDP)~\cite{Bernstein02:Complexity} have been proposed for reasoning in open environments~\cite{Chandrasekaran:Open,Cohen:OpenDecPOMDPs}.  The I-POMDP-Lite (as a relative of the more general I-POMDP) differs from the Dec-POMDP in several important ways.  First, the I-POMDP-Lite makes no assumptions that a subject agent observes the actual actions taken by other agents during operation in the environment, nor the observations they receive.  Indeed, each agent can reason in a vacuum, inferring the behaviors of others through changes to the shared environment state, whereas Dec-POMDPs assume that everyone's actions and observations are communicated or observed.  This makes the I-POMDP-Lite appropriate for a broader range of cooperative (and even competitive) environments, including the wildfire suppression domain where agents might not be able to pre-coordinate their behaviors, directly observe other agents obscured by smoke and fire, nor share information during operation due to damaged communication channels.

\subsection{Frame-Action Anonymity}

In many domains, it does not matter \emph{which} agents perform  actions, rather \emph{how many} agents of each frame perform each action. By relaxing agent identities,  multiple joint actions may produce the same action counts, and, in turn, equivalent state transitions, observations, and rewards. For example, in wildfire suppression, agents of different frames (e.g., ground firefighters, helicopters) extinguish fires at different rates.  But, each agent of the same frame contributes the same amount of fire suppression. Therefore, any $k$ agents of the same frame identically impact the fire. Such domains exhibit  \textbf{frame-action anonymity}~\cite{Sonu:2017}. 

Formally, let \textbf{configuration} $C = \left<n_{a_1, \theta_1}, ..., n_{a_{|A|}, \theta_{|\Theta|}} \right>$, where $n_{a_1, \theta_1}$ is the number of agents with frame $\theta_1$ performing action $a_1$ in a joint action tuple $(a_i, \ba_{-i})$. Histories with configurations (rather than joint actions) are now sufficient statistics \cite{Sonu:2017} and the representation of $T_i, O_i$, and $R_i$ can be greatly compacted since the number of possible configurations $|\mathcal{C}| \propto \binom{|Ag| + m + 1}{m + 1}$ is polynomial in the number of agents and thus much fewer than the exponential number of possible joint actions $m^{|Ag|}$, where $m = \max \{|A_i|, |A_j|, \ldots, |A_z|\}$ (c.f., Figure 1 below for comparisons of the number of configurations vs joint actions in our experiments).  Therefore, configurations enabled by frame-action anonymity improve scalability of individual planning in many-agent environments, as shown in the context of I-POMDPs~\cite{Sonu:2017}.  

Notably, in decision-theoretic planning, configurations are random variables since agents do not directly observe the actions chosen by other agents, but instead the likelihoods of different configurations (i.e., counts of actions per frame) are estimated based on the policies $\pi_{-i, l-1}$ solved for each other agent's mental model $M_{j, l-1}, \cdots, M_{z, l-1}$.

\section{Many-Agent Planning under Openness} 

In this paper, we limit our attention to systems where agents may leave the environment at any time and possibly reenter, but new agents do not enter the system. Still, this brings unique conceptual and computational challenges.

\subsection{Challenges of Open Agent Systems}

In open systems, individual planning is complicated by the need of each agent to track which other agents are currently present in the system and to reason about the actions of present agents only. In wildfire suppression, each firefighter must know how many others are currently unavailable due to recharging their suppressant, so as to focus on the behaviors of those currently fighting the fires.

Let $N(i) \subseteq Ag$ be the neighborhood of subject agent $i$, defined as the set of other agents with which $i$ can interact in the environment and that affect its transition, observation, and reward functions (e.g., the other firefighting robots that can suppress the same wildfires as subject agent $i$).  A naive way of tracking the presence of agents in open environments involves each agent $i$ maintaining an additional state variable $P_j$ for each neighbor $j \in N(i)$, where $P_j \in \{\text{present}, \text{absent}\}$. This increases the state space by a factor of $2^{|N(i)|}$ -- an exponential increase in the size of the agent's state space! To illustrate, modeling the presence or absence of 40 neighbors in this way increases the state space by a factor of $2^{40} \gg 100$ billion states. To avoid this, we place $P_j$ as an \emph{internal state variable within a mental model} $M_{j, l-1}$ attributed by subject agent $i$ to neighbor $j$.  Consequently, the overall increase in the size of the problem is \emph{linear} in the number of agents (scaled by the number of models ascribed to each agent), which promotes scalability.

Nonetheless, even a linear increase in the problem size poses challenges for planning.  In offline planning methods such as the interactive point-based value iteration~\cite{Chandrasekaran:Open} and interactive expectation-maximization~\cite{Qu:IEM}, the subject agent must still consider each neighbor's current presence and future transition, which adds at least a polynomial coefficient $O(|N(i)|^2)$ to the complexity of planning.  Likewise, in online planning approaches such as Monte Carlo tree search algorithms for POMDPs and I-POMDPs~\cite{Claes:Warehouse,Hula:MCTS}, the subject agent must spend time simulating the transitions of each neighbor when sampling an environment outcome, resulting in fewer sampled trajectories in the tree and thus lower quality approximations of the value function $V$.  Overall, these challenges motivate further steps to alleviate the impact of openness on computational complexity.

\subsection{Selectively Modeling Neighbors}

It is well known that surveyors and pollsters do not model every target person individually to understand their  attitudes and behaviors. Instead, the collective attitudes and behaviors of the whole are estimated by sampling those of a few, then extrapolating to everyone.  Bounds on the approximation error of this approach can be established through statistical analysis and random sampling~\cite{Neyman:Sampling,Frankel:History,Lohr:Sampling}.

Adapting this methodology to open, many-agent settings, we maintain mental models for a small subset of an agent's neighbors, then extrapolate their predicted behaviors to the entire neighborhood (with bounded error).  In the wildfire example, each firefighter models some of the firefighters with which it interacts, then uses their (fewer) predicted behaviors to estimate how many of the full set of firefighters are recharging or choosing to fight each fire.  By explicitly modeling some neighbors only, we reduce the reasoning necessary for considering changes to every neighbors' presence. Of course, the agent's planning process no longer obtains a policy for \emph{every} neighbor and thereby estimate the distribution of configurations necessary to estimate the state transitions, observations, and rewards (inside the $T_i^{a_i, o_i}(s', o_i | b_i, a_i)$ of Eq.~\ref{eqn:Q} and product in Eq.~\ref{eqn:rho}).  To mitigate this trade off, we use an alternative way to estimate configuration likelihoods when modeling a subset of the subject agent's neighborhood.  

\subsubsection{Estimating configuration probabilities}

Recall that during planning, configurations $C$ are random variables whose probability distributions typically depend on the policies obtained by the subject agent for its neighbors.  Alternately, since $C = \left<n_{a_1, \theta_1}, ..., n_{a_{|A|}, \theta_{|\Theta|}} \right>$ is composed of the counts of the number of agents of each frame performing each action, the distribution over $C$ can also be represented as a {\em multinomial distribution}, parameterized as
\begin{small}
\begin{align*}
P(C | s^t, M^t) \sim Multi(|N(i)|, \{p_{a_1, \theta_1, N(i)}, \ldots,  p_{a_{|A|}, \theta_{|\Theta|}, N(i)}\})
\end{align*}
\end{small}
\noindent where $p_{a, \theta, N(i)}$ is the probability that neighbors in $N(i)$ of frame $\theta$  will perform action $a$. We may view the configuration $C$ as the concatenation of several multinomial variables, $C_\theta = \left<n_{a_1, \theta}, n_{a_2, \theta}, \ldots, n_{a_{|A|}, \theta} \right>$, one for each frame $\theta$.  This allows us to model the multinomial distribution over $C$ using separate multinomial distributions for each $C_\theta$,
\begin{small}
\begin{align}
P(C_\theta | s^t, M^t) \sim Multi(|N_\theta(i)|, \{p_{a_1, N_\theta(i)}, \ldots, p_{a_{|A|}, N_\theta(i)}\})
\label{eqn:multitheta}
\end{align}
\end{small}
\noindent where $N_\theta(i) = \{j \in N(i) | j \text{ has frame } \theta\}$ is the set of $i$'s neighbors with frame $\theta$.  

If an agent chooses to model a proper subset of its neighbors $\hat{N_\theta}(i)$ $\subset$ $N_\theta(i)$, it can still estimate for each action the $p_{a, N_\theta(i)}$ values that parameterize the multinomial distribution $P(C_\theta | s^t, M^t)$.  Let $\hat{n}_{\pi(s^t)=a, \hat{N}_\theta(i)}$ be the number of agents in $\hat{N}_\theta(i)$ predicted to perform action $a$ at current state $s^t$ based on the policy obtained by solving their Nested-MDP model and their presence variable $P_j \in m_{j, l-1}$.  Then
\begin{align}
\hat{p}_{a, \hat{N}_\theta(i)} = \frac{\hat{n}_{\pi(s^t)=a, \hat{N}_\theta(i)}}{|\hat{N}_\theta(i)|} 
\label{eqn:p} 
\end{align}
is how likely an arbitrarily modeled neighbor of frame $\theta$ will perform action $a$.

Proportions $\hat{p}_{a, \hat{N}(i)}$ estimated using Eq.~\ref{eqn:p} allow us to estimate the underlying multinomial distribution over configuration likelihoods (Eq.~\ref{eqn:multitheta}).  We can then estimate the probability of a given configuration $C$ using the probability mass function of the modeled multinomial distribution.

\subsubsection{How many neighbors to model?}

Equations~\ref{eqn:multitheta} and \ref{eqn:p} are crucial in that they determine how we extrapolate the behaviors of a few modeled agents to the entire neighborhood \emph{without requiring explicit mental models and policies for every neighbor}.  At the same time, this approach approximates  the true distribution of configurations, which naturally raises questions such as how many neighbors should a subject agent model and how does the number of neighbors impact the error in the estimated configuration probabilities based on extrapolation.  We relate these two questions in the following theorem and corollary.  As Hoang and Low (\citeyear{Hoang13:Interactive}) previously established bounds on the error introduced by assuming that neighbors reason about others using a Nested-MDP, we focus our attention on errors introduced by modeling a subset of the neighbors.

Let the acceptable worst-case error in each of the $\hat{p}_{a, \hat{N}_{\theta}(i)}$ proportions in Eqs.~\ref{eqn:multitheta} and \ref{eqn:p} be $e_{\hat{p}}$.  We refer to this henceforth as the extrapolation error.  From statistical analysis~\cite{Lohr:Sampling}, we establish the minimum number of neighbors which an agent needs to explicitly model so that the extrapolation error does not exceed the given bound: 

\begin{thm}[Number of modeled neighbors] Let $N_\theta(i)$ be a neighborhood of agents with frame $\theta$ and whose size is $N$, $e_{\hat{p}}$ be a desired bound on extrapolation error, $(1-\alpha)$ be a statistical confidence level, and $t_{n-1, \frac{\alpha}{2}}$ come from the Student's t-distribution with $n-1$ d.o.f.  Then if agent $i$ models 
\begin{align}
n_\theta = |\hat{N_\theta}(i)| \ge \frac{N \left(\frac{t_{n-1, \frac{\alpha}{2}}}{2e_{\hat{p}}}\right)^2}{N - 1 + \left(\frac{t_{n-1, \frac{\alpha}{2}}}{2e_{\hat{p}}}\right)^2}
\label{eqn:pnfpc2} 
\end{align}
\noindent neighbors, then it will be confident at the $(1-\alpha)$ level that $\hat{p}_{a, \hat{N_\theta}(i)}$ for each action $a$  will be within $e_{\hat{p}}$ of the true proportions of all agents choosing action $a$.
 \label{thm:modeling2}
\end{thm}

The proof of this theorem is provided in Appendix A of the supplementary material.  To illustrate, let $N = 50$.  To ensure that the agent's estimated $\hat{p}_{a, \hat{N}_\theta(i)}$ extrapolates to within $e_{\hat{p}} = 0.1$ of the true proportion for the entire neighborhood with frame $\theta$ with 95\% confidence, we need model at least 34 neighbors. This drops to 18 if instead $e_{\hat{p}}$ = 0.2. 

If the agent indeed models at least the prescribed number of other agents from Theorem~\ref{thm:modeling}, the error in each configuration's probability estimate is bound by the value $\epsilon_{P(C)}$ given below (proof in Appendix B of the supplementary material).

\begin{cor}[Error bound on configuration probability ]
Let $n_{\theta}$ be the number of neighbors given by Theorem 1 (for a given confidence level $1-\alpha$) that subject agent $i$ chooses to model from its neighborhood $N_{\theta}(i)$ for each $\theta \in \Theta$, and let $\hat{p}_{a, \theta}$ be the resulting estimated proportions of agents within those neighborhoods that will choose action $a$, given state $s$ and mental models $M$.  Then the estimated probability $P(C | s^t, M^t)$ that the entire neighborhood will exhibit the configuration $C$ has error $\epsilon_{P(C)}$ due to modeling $n_\theta$ agents only, which is less than:
\begin{align}
\small
&|P^*(C | s^t, M^t) - P(C | s^t, M^t)| = \epsilon_{P(C)}\nonumber\\
&< \frac{\prod_\theta{|N_\theta(i)|!}}{\prod_{a, \theta}{C(a, \theta)!}}
~\left[ \prod_{a, \theta}{\left(\hat{p}_{a, \theta}+e_{\hat{p}}\right)^{C(a, \theta)}} - \prod_{a, \theta}{\hat{p}_{a, \theta}^{C(a, \theta)}}\right]
\label{eqn:errorPC} 
\end{align}
where $P^*(C | s^t, M^t)$ denotes the \emph{true} likelihood that configuration $C$ will result from state $s^t$ and mental model $M^t$.
\label{cor:errorPC}
\end{cor}

\subsubsection{Regret Bounds}

As modeling a subset of neighbors produces a bounded approximation of the true configuration probabilities considered during planning, the resulting value function $V$ will also be an approximation. Next, we establish that any error in the approximate value function (and hence the discounted cumulative rewards actually earned by the agent following the corresponding policy $\pi_i$) is also bounded.  Here, $J_{i, k}$ represents the actual discounted cumulative rewards the agent would earn following $\pi_i$ for $k$ time steps, whereas $V^*_{i, k}$ is the optimal value function.

\begin{thm}(Regret bound).
Maximum regret that agent $i$ incurs $\left\| V_{i, k}^* - J_{i, k} \right\|_\infty$ from following a $k$-horizon optimal policy $\pi_i$ (obtained by solving the many-agent I-POMDP-Lite) due to the approximate likelihoods of other agents' configurations $P(C | s^t, M^t)$ is bounded from above:

\begin{align}
\begin{split}
\left\| V_{i, k}^* - J_{i, k} \right\|_\infty & \le 2 \epsilon_{P(C)}\cdot |\mathcal{C}| \cdot R_{max} \\ & \times \left[\gamma^{k-1} + \frac{1}{1-\gamma} \left(1 + 3\gamma\frac{|\Omega_i|}{1 - \gamma}\right) \right]
\end{split}
\label{eqn:regretbound}
\end{align}
\label{thm:regretbound}
\end{thm}

Notice that the bound is linear in the error $\epsilon_{P(C)}$ in the agent's estimation of configuration likelihoods caused by modeling some neighbors only. The proof is provided in Appendix C of the supplementary material. Though the resulting value function from planning might not be exact, the approximation error is upper bounded and is proportional to only $\sqrt{n_\theta}$ (in the worst case) due to the fact that $\epsilon_{P(C)}$ is at worst linear in $e_{\hat{p}}$ (c.f., Eq.~\ref{eqn:errorPC} since $e_{\hat{p}} < 1$), and $e_{\hat{p}}$ is proportional to $\sqrt{n_\theta}$ (c.f., Eq.~\ref{eqn:pnfpc}).

\subsubsection{Which neighbors to explicitly model?}

After determining the number of agents to model for each frame-action pair, we must select which neighbors will be explicitly modeled, $\hat{N}_\theta(i)$. This selection is performed only when the agent joins the environment. Thus, it does not add to the computational complexity of planning.

To decide which neighbors to model, we first note that we can decompose the agent's neighborhood $N_\theta(i)$ into (potentially overlapping) sets based on the actions that the neighbors {\em can} perform: $N_{\theta,a}(i) = \{j \in N_\theta(i) | j \text{ can perform action } a\}$.  Each $N_{\theta, a}(i)$ provides a sample set from which we select neighbors in order to estimate $\hat{p}_{a, \hat{N}_\theta(i)}$.  To construct $\hat{N_\theta}(i)$, we randomly sample $n$ agents (Eq.~\ref{eqn:pnfpc}) for each $(\theta, a)$ pair.

Second, many agents may appear in more than one such sample set because agents can perform more than one action.  Thus, once a neighbor $j$ has been added to $\hat{N_\theta}(i)$, it contributes to the $n$ needed for each $(\theta, a)$ pair corresponding to the actions that agent $j$ can perform.  If sampling neighbors starts with the largest $N_{\theta, a}$ and proceeds to the smallest, then agents that have the greater number of possible actions (e.g., can fight the most fires) will have the largest total chance of being sampled for modeling since they appear in most subsets $N_{\theta, a}(i)$.  This property is beneficial as such agents are expected to be the \emph{most influential agents} in the system, as well as the \emph{most necessary to explicitly model} because their reasoning will be more complex than other agents, and thus their behaviors are more difficult to predict.

\section{Many-Agent MCTS for Open Systems} 

Based on our approach for modeling a subset of an agent's neighbors, we next extend the popular POMCP algorithm \cite{Silver10:POMCP} to create a MCTS algorithm for many-agent open environments.  

\subsubsection{Single-agent MCTS}
A popular approach to quickly approximate the value function and policy of a POMDP is Monte Carlo tree search (MCTS)~\cite{Ross08:Online,Silver10:POMCP,Somani13:DESPOT}.  In MCTS, the agent iteratively constructs an AND-OR tree representing the possible beliefs that can be reached by following sequences of actions, and obtains the expected discounted utility of performing those action sequences by simulating future rewards.  MCTS performs online planning for the beliefs that an agent actually encounters, avoiding the need for computationally-expensive offline planning and scales well to large state spaces due to MC sampling of future state-action-reward trajectories.  Until the time budget is met, MCTS revises estimated $Q(b, a)$ values along the trajectories and extending the tree at the leaf using rollout.

\algrenewcommand\algorithmicindent{0.4em}
\begin{algorithm}[!ht]
\small
\caption{I-POMCP$_\mathcal{O}$: Open Many-Agent MCTS}
\textbf{Note:} $T$ is the tree (initially empty), $p$ is a path from the root of the tree (with $p = \emptyset$ signifying the root), $B_{p}$ is the particle filter signifying the set of state-model pairs encountered at the node at $p$ in the tree, $PF$ is the root particle filter, $N$ is count of the number of visits to each node in the tree initialized to some constant $\nu \ge 0$, $Q$ is the Q function initialized to 0, $c$ a constant from UCB-1.

\begin{algorithmic}[1]
\Procedure{\ipomdp{}-MCTS}{$PF, \tau$}
\State $time \leftarrow 0$
\While{$time < \tau$}
\State $s^0, M^0 \leftarrow SampleParticle\left(PF\right)$
\State UpdateTree$\left(s^0, M^0, 0, \emptyset\right)$
\State{Increment $time$}
\EndWhile
\State return $\underset{a \in A_i}{\operatorname{argmax}} \text{ } Q(\emptyset, a)$
\EndProcedure
\end{algorithmic}

\begin{algorithmic}[1]
\Procedure{UpdateTree}{$s^t, M^t, t, p$}
\If {$t \ge H$}
\State return 0
\EndIf
\State $B_{p} \leftarrow B_{p} \cup \{\left(s^t, M^t\right)\}$ 
\If {$p \notin T$}
\State $T \leftarrow T + \text{leafnode}(p)$
\State return Rollout$\left(s^t, M^t, t\right)$
\Else
\State $C^t$ $\leftarrow$ SampleConfiguration$\left(s^t, M^t\right)$
\State $a_i^t \leftarrow \underset{a \in A_i}{\operatorname{argmax}} \text{ } Q\left(p, a\right) + c \sqrt{\left(\log N_{p}\right) / N_{p \mapsto a}}$
\State $s^{t+1}, M^{t+1}, o_i^t, r_i^t$ $\leftarrow$ Simulate$\left(s^t, M^t, a_i^t, C^t\right)$
\State $N_{p} \leftarrow 1 + N_{p}$
\State $N_{p \mapsto a_i^t} \leftarrow 1 + N_{p \mapsto a_i^t}$
\State $p' \leftarrow p + \left(a_i^t, o_i^t \right)$ \State $R \leftarrow r_i^t + \gamma \cdot$UpdateTree$\left(s^{t+1}, M^{t+1}, t+1, p'\right)$
\State $Q(p, a_i^t) \leftarrow Q(p, a_i^t) + R - Q(p, a_i^t) / N_{p \mapsto a_i^t}$ 
\State return $R$
\EndIf
\EndProcedure
\end{algorithmic}

\begin{algorithmic}[1]
\Procedure{Rollout}{$s^t, M^t, t$}
\State $R \leftarrow 0, t' \leftarrow t$
\While{$t < H$}
\State $C^{t}$ $\leftarrow$ SampleConfiguration$(s^{t}, M^{t})$
\State $a_i^{t}$ $\leftarrow$ SampleAction$(A_i)$
\State $s^{t+1}, M^{t+1}, o_i^{t}, r_i^{t}$ $\leftarrow$ Simulate$(s^{t}, M^{t}, a_i^{t}, C^{t})$
\State $R \leftarrow R + \gamma^{t - t'} \cdot r_i^t$, $t \leftarrow t + 1$
\EndWhile
\State return $R$
\EndProcedure
\end{algorithmic}

\begin{algorithmic}[1]
\Procedure{SampleConfiguration}{$s^t, M^t$}
\State $C(a, \theta) \leftarrow 0$, $\hat{n}_{\pi(s^t)=a, \hat{N}_\theta(i)} \leftarrow 0 \text{ } ~~~\forall a, \theta$
\For{$\M_{j,l-1} \in M^t$}
\State $a \sim \pi_{j, l-1}(s^t)$
\State $\hat{n}_{\pi(s^t)=a, \hat{N}_{\theta_j}(i)} \leftarrow \hat{n}_{\pi(s^t)=a, \hat{N}_{\theta_j}(i)} +  1$
\EndFor
\For{$\theta \in \Theta$}
\For{$a \in A$}
\State $\hat{p}_{a, \hat{N(i)}} \leftarrow \hat{n}_{\pi(s^t)=a, \hat{N}_\theta(i)} ~/~ |\hat{N}_\theta(i)|$
\EndFor
\For{$j \in N_\theta(i)$}
\State $a \sim Cat(\hat{p}_{a_1, \hat{N}_\theta}, \hat{p}_{a_2, \hat{N}_\theta}, \ldots, \hat{p}_{a_{|A|}, \hat{N}_\theta})$
\State $C(a, \theta) \leftarrow C(a, \theta) + 1$
\EndFor
\EndFor
\State return $C$
\EndProcedure
\end{algorithmic}
\label{alg:many-agent MCTS}
\end{algorithm}

\subsubsection{Many-agent MCTS}

We generalize MCTS to open, many-agent settings in the context of the I-POMDP-Lite framework (also applicable to I-POMDPs), exploiting both frame-action anonymity and selective neighbor modeling to further enhance scalability.  Our algorithm, shown in Algorithm~\ref{alg:many-agent MCTS}, is \textbf{I-POMCP$_\mathcal{O}$}.  It considers agent openness in ($i$) {\sf Simulate}, which additionally simulates the agents leaving and reentering (in $M$), and in ($ii$) {\sf  SampleConfiguration} when obtaining configuration probabilities.

Our I-POMCP$_\mathcal{O}$ algorithm differs from the classic single agent POMCP \cite{Silver10:POMCP} in the following ways.  First, the agent's rewards $R_i$, state transitions $T_i$, and observations $O_i$ depend on not only its own actions and the environment state, but also the configuration of actions by the other agents.  Thus, before the agent can simulate the environment on line 6 of the {\sf Rollout} and line 11 of the {\sf UpdateTree} procedures, the agent must sample a configuration $C^t$ of its neighbors' actions.  This addition to standard POMCP occurs on lines 4 and 9 of the two procedures.  

Second, the process for sampling a configuration exploits the multinomial probability distribution $P(C^t | s^t, M^t)$ determined from the models of a select subset of the agent's neighbors in the {\sf SampleConfiguration} procedure, so that the agent need not model every neighbor.  Since only some mental models are consulted, there is opportunity to save computation time in this procedure, as opposed to modeling every neighbor\footnote{Provided that sampling an action for each neighbor is less computationally expensive than interacting with an actual policy.}.  Moreover, since only some mental models are needed by this procedure, the {\sf Simulate} procedure modeling the environment (called on lines 6 and 11 of {\sf Rollout} and {\sf UpdateTree}, respectively) needs to process fewer mental model transitions, saving additional computational time.  Overall, this should require less time per trajectory through the AND-OR tree in each call of {\sf UpdateTree} from the root of the tree, and thus more total trajectories sampled and closer approximations of the true value function.  

This algorithm significantly differs from previous MCTS algorithms for multiagent POMDPs.  Hula et al. (\citeyear{Hula:MCTS}) first adapted POMCP to an I-POMDP setting.  However, they made several assumptions in their adaptation to their particular problem: the presence of only two agents, agents' actions occur sequentially rather than simultaneously, and agents directly observe each other actions.   Their approach would not scale to many-agent environments as it would have computational complexity that is exponential in the number of agents.  Our approach applies to more general settings, assuming only frame-action anonymity, and exhibits complexity that is \emph{linear} in the number of agents: $O(IHN)$ where $I$ is the number of sampled MC trajectories, $H$ is the horizon, and $N$ is the total number of modeled agents.  Amato and Oliehoek (\citeyear{Amato:Scalable}) and Best et al. (\citeyear{Best:Dec-MCTS}) proposed similar algorithms for Dec-POMDPs that scale to more than two agents.  However, our algorithm enables planning in settings when the joint actions of other agents are unknown and must be estimated through individual models, as well as incorporates frame-action anonymity and selective neighbor modeling for improved scalability.  Claes et al. (\citeyear{Claes:Warehouse}) proposed an algorithm that simplified the models of neighbor behavior using domain heuristics to enhance scalability, as well as social laws to help promote cooperation.  Here, however, we consider more general models of neighbor behavior that will work across domains without specific heuristics.

\section{Experiments}

We evaluate implementations\footnote{Available at https://github.com/OberlinAI/ScalableOASYS} of I-POMCP$_\mathcal{O}$ on the wildfire suppression problem. Chandrasekaran et al. (\citeyear{Chandrasekaran:Open}) used this domain to study planning in \emph{small} open agent environments; we extend it to include more fires and significantly more firefighters and corresponding joint actions.

\subsubsection{Setups} 
\begin{figure}
\centering
\includegraphics[width=.95\columnwidth]{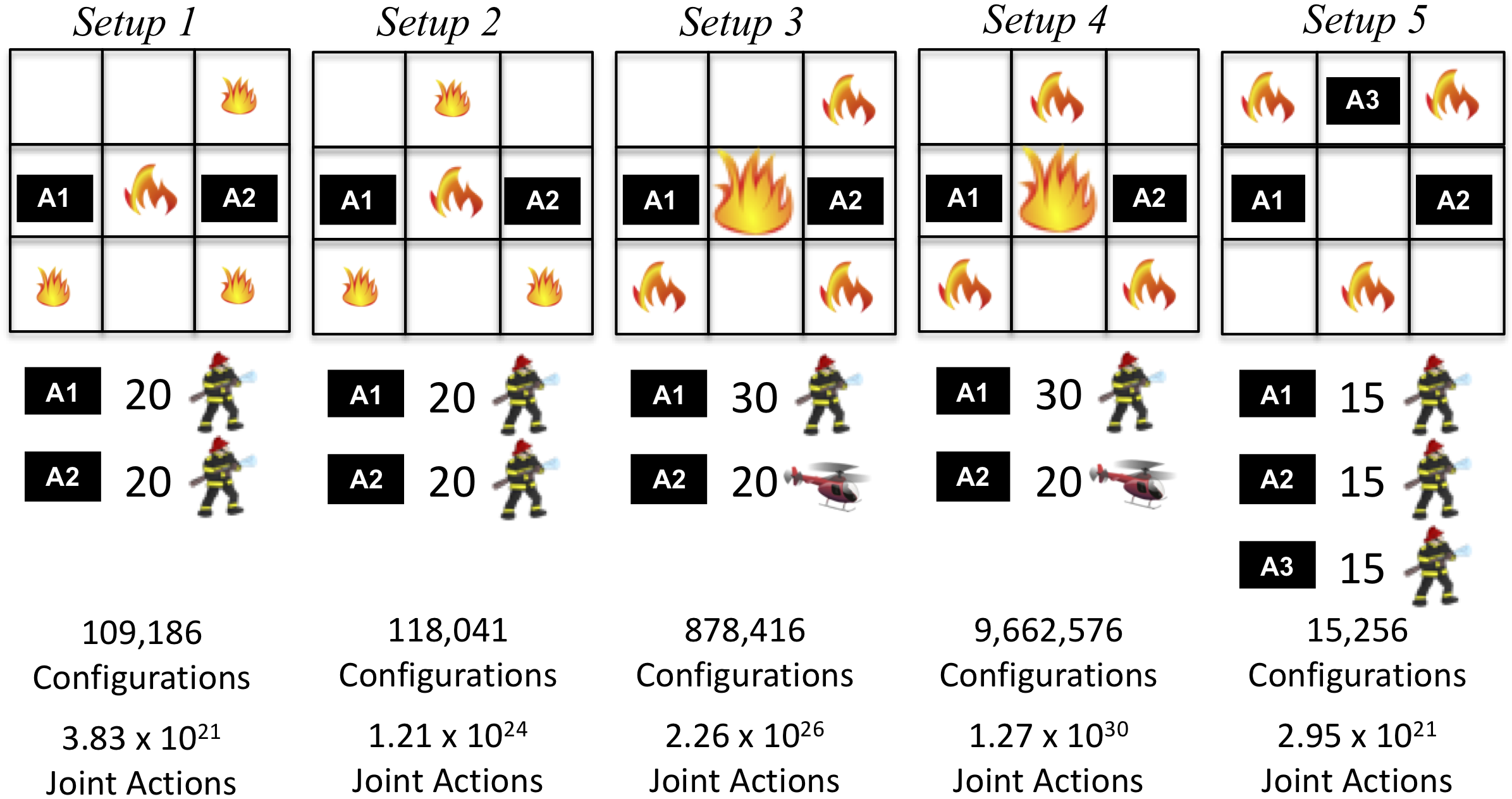}
\caption{Our environment setups involve a varying number of fires, fire intensities and positions, agents and their types.}
\label{fig:setups}
\end{figure}

Agents are tasked with putting out fires of different sizes in the absence of inter-agent communication and prior coordination. Small, large, and huge fires require at least 10, 20, and 30 agents to act together to reduce their intensity, respectively. The spread of fires is modeled on the dynamics of real wildfires~\cite{Boychuk09:Fire,Ure15:Fire}. Agents have limited amounts of suppressants that stochastically transition from full to half-full to empty, and back to full on recharging during which they temporarily leave the environment for an average of two steps, then rejoin when full. Each agent can use its suppressant on a fire in an adjacent location or take a NOOP action; the latter also represents actions to take while recharging suppressant. Agents earn shared rewards of 20, 40, and 60 for putting out small, large, and huge fires.  They receive shared penalties of 1 when a location burns out and individual penalties of 100 for either fighting a non-existent fire or choosing a non-NOOP action without suppressant. Figure~\ref{fig:setups} illustrates the environments used in our experiments. Notice that our agents are ground firefighters and helicopters, whose frames differ: helicopters are twice as effective in their firefighting ability.  Though several agents share the same location, they start with different suppressant levels. As such, they plan and behave differently rather than as one collective agent.

Though in seemingly small grids, these decision problems are non-trivial and substantially complex, especially compared to the prior literature -- with 40-50 agents, the number of joint actions ranges from $2.95 \times 10^{21}$ to $1.27 \times 10^{30}$.  Even with frame-action anonymity, there are 15,256 to 9,662,576 configurations; thus, these setups are intractable for prior planning algorithms.  We intentionally do not consider smaller domains for which existing algorithms are tractable (e.g., with only 5 agents \cite{Chandrasekaran:Open}), as there would be no need for modeling only a few neighbors (Thm.~\ref{thm:modeling} would model every neighbor $\forall e_{\hat{p}} < 1$).

\subsubsection{Metrics and baselines} Agent performance is measured by: ($i$) the average rewards earned by agents, ($ii$) the average number of fire locations put out, and ($iii$) the average amount of suppressant used. These measures evaluate how how well agents \emph{optimize} their planning problems, and how \emph{effectively} and \emph{efficiently} they put out fires, respectively.

To demonstrate both the efficacy and impact of our strategy of intelligently modeling only a subset of neighbors, we consider variants of \textbf{I-POMCP}$_\mathcal{O}$, each planning at level $l = 2$, that use different $e_{\hat{p}} \in [0, 0.3]$ with Theorem~\ref{thm:modeling} providing the numbers of neighbors to model (with $\alpha = 0.05$).  Here, $e_{\hat{p}}=0$ represents I-POMCP$_\mathcal{O}$ modeling every neighbor and also provides an upper bound on how the algorithm of Hula et. al (\citeyear{Hula:MCTS}) would perform if generalized to many-agent environments and Amato and Oliehoek (\citeyear{Amato:Scalable}) if extended to the I-POMDP-Lite with frame-action anonymity.

We compare the performance of our approach with two baselines (using Kruskal-Wallis followed by post-hoc pairwise Mann-Whitney for significance testing). First, \textbf{NestedMDP} represents an agent using an adaptation of VI to perform $\epsilon$-bounded planning over configurations at level $l = 1$.  This is a strong baseline, which serves as both a scalable, fully observable approximation that can tractably solve our decision problems, as well as a comparison to the models that agent $i$ assumes its neighbors follow.  Second, \textbf{Heuristic} is an agent that randomly chooses an adjacent fire to fight if one exists and the agent has suppressant, else it takes a NOOP action. It represents semi-intelligent behavior in complex environments (similar to domain heuristics in multirobot dynamic task allocation, e.g., \cite{Lerman:DynamicMRTA}).  Of note, traditional I-POMDP solvers do not scale to many-agent environments.  Indeed, we ran the previous method, interactive point-based value iteration~\cite{Chandrasekaran:Open}, on Setup 1 with only 10 total agents; after almost a week of planning, it did not produce a usable policy.

\begin{figure}
\centering
\includegraphics[width=.95\columnwidth]{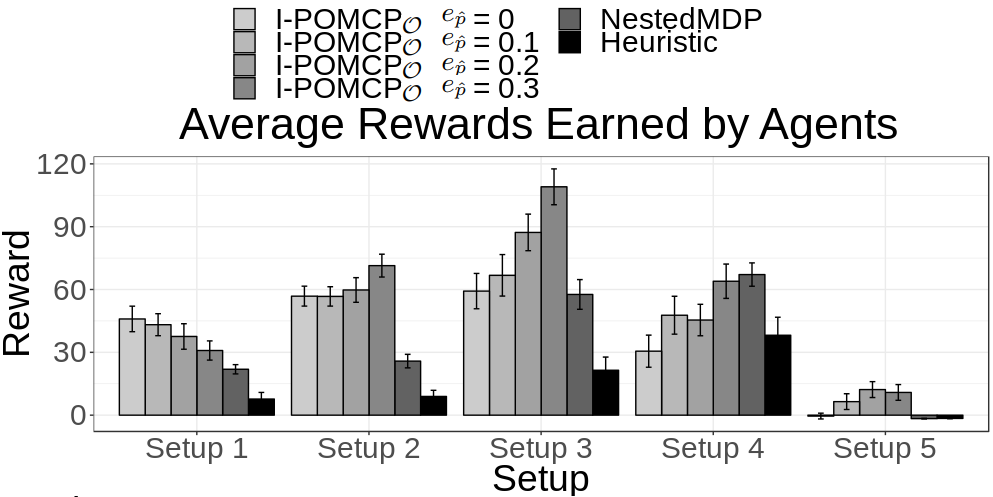}
\caption{Total reward earned per agent averaged across 100 runs. Error bars represent 95\% CIs. I-POMCP$_\mathcal{O}$ planned for 5 sec. in Setups 1-3 and 5, and for 20 sec. in Setup 4.}
\label{fig:rewards}
\end{figure}

\subsubsection{Results} We observe from Fig.~\ref{fig:rewards} that I-POMCP$_\mathcal{O}$ improved on NestedMDP and Heuristic by achieving significantly greater rewards in Setups 1-3 and 5, and statistically equivalent reward to NestedMDP in Setup 4. Further investigations (c.f., Figs. 4 and 5 in Appendix D of the supplementary material) reveal that I-POMCP$_\mathcal{O}$ agents used slightly more suppressant to put out fires at significantly more locations than NestedMDP (true in all setups). NestedMDP agents earned comparable rewards in Setup 4 by concentrating all of their efforts on a single shared fire, whereas I-POMCP$_\mathcal{O}$ agents spread their resources across the most complex environment.  Interestingly, in Setup 5, which had the most shared fires, NestedMDP never put out a fire. This was due to the symmetry in the problem causing NestedMDP to behave randomly, whereas I-POMCP$_\mathcal{O}$ was more successful.

Comparing the different settings of I-POMCP$_\mathcal{O}$, increasing $e_{\hat{p}}$ so that fewer neighbors were modeled led to \emph{increased performance} in the more complicated Setups 2-5.  Moreover, agents were also \emph{more efficient} with their suppressants, using less resources to put out more fires as they modeled fewer neighbors. These results were obtained because MCTS could then run a greater number of trajectories in the fixed planning time $\tau$, in essence trading off modeling accuracy for better value estimates.  Furthermore, in Setup 5, modeling only some neighbors biased agents to favor one shared fire over another, breaking the complicated symmetry of the problem. On the other hand, modeling fewer neighbors reduced performance in only the simplest Setup 1, which had only one shared fire and one frame of agents.  In this setup, even when modeling all agents, enough trajectories were sampled to adequately estimate the value function.

\section{Concluding Remarks}

Real-world domains often exhibit agent openness, where agents may leave and then return. Scaling in the number of agents in such domains continues to remain a challenge for planning in realistic multiagent environments.  Our method -- consisting of an approach to selectively model neighbors that is new to planning and a generalized MCTS algorithm for many-agent settings -- models the presence or absence of agents as it predicts their behaviors. Our key insight for scaling is that we may explicitly model just a few agents and extrapolate the model predictions to others with provable error bounds. This improves on baselines as we demonstrated on a spectrum of firefighting scenarios with an order of magnitude more agents than previously considered in open environments. The idea of extrapolating modeling is new and not limited to open agent settings. It could be combined in future work with other approximations that promote scalability.  

\section{Acknowledgments}
This research was supported in part by NSF grant \#1909513 (to AE, PD, and LS).  Experiments were conducted on Oberlin College's SCIURus cluster (NSF Grant \#1427949).

\bibliographystyle{aaai}
\bibliography{AAAI-EckA.6382}

\newpage
\appendix
\setcounter{secnumdepth}{1} 
\setcounter{equation}{5}

\section{Proof for Theorem 1}
\setcounter{thm}{0}
\begin{thm}[Number of modeled neighbors] Let $N_\theta(i)$ be a neighborhood of agents with frame $\theta$ and whose size is $N$, $e_{\hat{p}}$ be a desired bound on extrapolation error, $(1-\alpha)$ be a statistical confidence level, and $t_{n-1, \frac{\alpha}{2}}$ come from the Student's t-distribution with $n-1$ d.o.f.  Then if agent $i$ models 
\begin{align}
n_\theta = |\hat{N_\theta}(i)| \ge \frac{N \left(\frac{t_{n-1, \frac{\alpha}{2}}}{2e_{\hat{p}}}\right)^2}{N - 1 + \left(\frac{t_{n-1, \frac{\alpha}{2}}}{2e_{\hat{p}}}\right)^2}
\label{eqn:pnfpc} 
\end{align}
\noindent neighbors, then it will be confident at the $(1-\alpha)$ level that $\hat{p}_{a, \hat{N_\theta}(i)}$ for each action $a$  will be within $e_{\hat{p}}$ of the true proportions of all agents choosing action $a$.
 \label{thm:modeling}
\end{thm}

\setcounter{equation}{8}
\paragraph{Proof} Let $\hat{p}$ denote $\hat{p}_{a, \hat{N}_\theta(i)}$ and $n$ denote $|\hat{N}_\theta(i)|$.  With statistical confidence $1-\alpha$, the true proportion of neighbors of frame $\theta$ who will choose action $a$ will lie within the range:

\begin{align}
\hat{p} \pm t_{n-1, \frac{\alpha}{2}} \sqrt{\frac{\hat{p}(1-\hat{p})}{n}}
\label{eqn:prange} 
\end{align}
However, the range in Eq.~\ref{eqn:prange} is rather loose as it assumes the neighborhood $N_\theta(i)$ is infinitely sized. For finite neighborhoods, we can tighten this range using the finite population correction~\cite{Lohr:Sampling}.  Let $N = |N_\theta(i)|$ be the size of the finite neighborhood. Then the error bound is:
\begin{align}
\hat{p} \pm t_{n-1, \frac{\alpha}{2}} \sqrt{\frac{\hat{p}(1-\hat{p})}{n}}\sqrt{\frac{N-n}{N-1}}
\label{eqn:prangefpc} 
\end{align}
So, with confidence $1-\alpha$ the error $e_{\hat{p}}$ of $\hat{p}$ is:
\begin{small}
\begin{align*}
e_{\hat{p}} \le t_{n-1, \frac{\alpha}{2}} \sqrt{\frac{\hat{p}(1-\hat{p})}{n}}\sqrt{\frac{N-n}{N-1}} \le t_{n-1, \frac{\alpha}{2}} \sqrt{\frac{0.5^2}{n}}\sqrt{\frac{N-n}{N-1}}.
\end{align*}
\end{small}
The last inequality results because the range in Eq.~\ref{eqn:prangefpc} attains maximal width when $\hat{p} = 0.5$.  Reorganizing this inequality yields Eq.~\ref{eqn:pnfpc}. $\square$

\setcounter{figure}{2}
Of note, Fig.~\ref{fig:samplesizes} demonstrates the relationship between $n$ and $N$ in Theorem 1 for different values of $e_{\hat{p}}$ for the standard $95\%$ significance level.

\begin{figure}[!ht]
\centering
\includegraphics[width=2.5in]{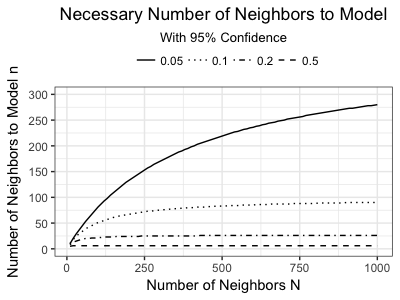}
\caption{Relationships between n and N}
\label{fig:samplesizes}
\end{figure}

\section{Proof for Corollary 1}


Note: in the following, we drop the timestep $t$ notation as our theoretical analysis does not depend on it, but instead generalizes to all time steps.

\setcounter{cor}{0}
\setcounter{equation}{6}
\begin{cor}[Error bound on configuration probability ]
Let $n_{\theta}$ be the number of neighbors given by Theorem 1 (for a given confidence level $1-\alpha$) that subject agent $i$ chooses to model from its neighborhood $N_{\theta}(i)$ for each $\theta \in \Theta$, and let $\hat{p}_{a, \theta}$ be the resulting estimated proportions of agents within those neighborhoods that will choose action $a$, given state $s$ and mental models $M$.  Then the estimated probability $P(C | s^t, M^t)$ that the entire neighborhood will exhibit the configuration $C$ has error $\epsilon_{P(C)}$ due to modeling $n_\theta$ agents only, which is less than:
\begin{align}
\small
&|P^*(C | s, M) - P(C | s, M)| = \epsilon_{P(C)}\nonumber\\
&< \frac{\prod_\theta{|N_\theta(i)|!}}{\prod_{a, \theta}{C(a, \theta)!}}
~\left[ \prod_{a, \theta}{\left(\hat{p}_{a, \theta}+e_{\hat{p}}\right)^{C(a, \theta)}} - \prod_{a, \theta}{\hat{p}_{a, \theta}^{C(a, \theta)}}\right]
\label{eqn:errorPC2} 
\end{align}
where $P^*(C | s, M)$ denotes the \emph{true} likelihood that configuration $C$ will result from state $s$ and mental model $M$.
\label{cor:errorPC2}
\end{cor}

\setcounter{equation}{10}
\paragraph{Proof} Recall from Section 4.2.1 that each configuration $C$ can be modeled as the concatenation of a separate $C_\theta$ for each $\theta \in \Theta$.  The likelihood of an arbitrary $C_\theta$ is given by the probability mass function of the multinomial distribution resulting from the $\hat{p}_{a, \theta}$ estimated from each neighborhood $N_\theta(i)$:
\begin{align}
P(C_\theta | s, M) = |N_\theta(i)|! \prod_{a, \theta}{\frac{\hat{p}_{a, \theta}^{C(a, \theta)}}{C(a, \theta)!}} 
\label{eqn:PCTheta} 
\end{align}

Thus, the probability of the fully concatenated configuration $C$ is given by:
\begin{align}
P(C | s, M) = \prod_\theta{|N_\theta(i)|!} \prod_{a}{\frac{\hat{p}_{a, \theta}^{C(a, \theta)}}{C(a, \theta)!}} 
\label{eqn:PC} 
\end{align}

Based on the choice of $n_{\theta}$ from Theorem 1, each $\hat{p}_{a, \theta}$ could \emph{over}-estimate the true proportions by at most $\epsilon_{\hat{p}}$ (with confidence $1 - \alpha$).  Indeed, if instead the true proportions of agents from each neighborhood were their lowest values from within their corresponding margins of errors, then the true probability mass function would be:

\begin{align}
\underline{P}(C | s, M) = \prod_\theta{|N_\theta(i)|!} \prod_{a}{\frac{\left(\hat{p}_{a, \theta} - \epsilon_{\hat{p}}\right)^{C(a, \theta)}}{C(a, \theta)!}} 
\label{eqn:PClower} 
\end{align}

and thus Eq.~\ref{eqn:PC} could over estimate the true probability of an arbitrary configuration $C$ by at most:

\begin{align}
\begin{split}
P(C | s, M) &- \underline{P}(C | s, M) {} 
\\ & = \prod_\theta{|N_\theta(i)|!} \prod_{a}{\frac{\hat{p}_{a, \theta}^{C(a, \theta)}}{C(a, \theta)!}} 
\\ &- \prod_\theta{|N_\theta(i)|!} \prod_{a}{\frac{\left(\hat{p}_{a, \theta} - \epsilon_{\hat{p}}\right)^{C(a, \theta)}}{C(a, \theta)!}}  \\
& = \frac{\prod_\theta{|N_\theta(i)|!}}{\prod_{a, \theta}{C(a, \theta)!}}\left[\prod_{a, \theta}{\hat{p}_{a, \theta}^{C(a, \theta)}} - \prod_{a, \theta}{\left(\hat{p}_{a, \theta}+\epsilon_{\hat{p}}\right)^{C(a, \theta)}}\right]
\end{split}
\label{eqn:PClowererror} 
\end{align}

Analogously, each $\hat{p}_{a, \theta}$ could \emph{under}-estimate the true proportions by at most $\epsilon_{\hat{p}}$.  Indeed, if instead the true proportions of agents from each neighborhood were their largest values from within their corresponding margins of errors, then the true probability mass function would be:

\begin{align}
\overline{P}(C | s, M) = \prod_\theta{|N_\theta(i)|!} \prod_{a}{\frac{\left(\hat{p}_{a, \theta} + \epsilon_{\hat{p}}\right)^{C(a, \theta)}}{C(a, \theta)!}} 
\label{eqn:PCupper} 
\end{align}

and thus Eq.~\ref{eqn:PC} could under estimate the true probability of an arbitrary configuration $C$ by at most:

\begin{align}
\begin{split}
\overline{P}(C | s, M) & - P(C | s, M)
\\ & = \prod_\theta{|N_\theta(i)|!} \prod_{a}{\frac{\left(\hat{p}_{a, \theta} + \epsilon_{\hat{p}}\right)^{C(a, \theta)}}{C(a, \theta)!}} 
\\ & - \prod_\theta{|N_\theta(i)|!} \prod_{a}{\frac{\hat{p}_{a, \theta}^{C(a, \theta)}}{C(a, \theta)!}} \\
& = \frac{\prod_\theta{|N_\theta(i)|!}}{\prod_{a, \theta}{C(a, \theta)!}}\left[\prod_{a, \theta}{\left(\hat{p}_{a, \theta}+\epsilon_{\hat{p}}\right)^{C(a, \theta)}} - \prod_{a, \theta}{\hat{p}_{a, \theta}^{C(a, \theta)}}\right]
\end{split}
\label{eqn:PCuppererror} 
\end{align}

Within the largest possible errors in Eq.~\ref{eqn:PClowererror} and Eq.~\ref{eqn:PCuppererror}, we note that for arbitrary action $a$ and frame $\theta$, we must have that:
\begin{align}
\left(\hat{p}_{a, \theta} + \epsilon_{\hat{p}}\right)^{C(a, \theta)} - \hat{p}_{a, \theta}^{C(a, \theta)} > \hat{p}_{a, \theta}^{C(a, \theta)} - \left(\hat{p}_{a, \theta} - \epsilon_{\hat{p}}\right)^{C(a, \theta)}
\end{align}

by the binomial expansion of the $\left(\hat{p}_{a, \theta} + \epsilon_{\hat{p}}\right)^{C(a, \theta)}$ and $\left(\hat{p}_{a, \theta} - \epsilon_{\hat{p}}\right)^{C(a, \theta)}$ terms.  Therefore, we find that:

\begin{align}
\overline{P}(C | s, M) - P(C | s, M) > P(C | s, M) - \underline{P}(C | s, M)
\label{eqn:PCcompareerror} 
\end{align}

Based on our definitions above of $\overline{P}(C | s, m)$ and $\underline{P}(C | s, M)$, we have that:
\begin{align}
\underline{P}(C | s, M) < P^*(C | s, M) < \overline{P}(C | s, M)
\label{eqn:PCboundtrue} 
\end{align}

where the strict inequalities come from the fact that the probabilities $\{\hat{p}_{a, \theta} - \epsilon_{\hat{p}} | \forall a \in A \}$ do not form a valid probability distribution because they sum to less than 1, so Eq. ~\ref{eqn:PClower} must be strictly less than $P^{*}(C | s, M)$.  Similarly, the probabilities $\{\hat{p}_{a, \theta} + \epsilon_{\hat{p}} | \forall a \in A \}$ do not form a valid probability distribution because they sum to more than 1, so Eq. ~\ref{eqn:PCupper} must be strictly greater than $P^{*}(C | s, M)$.

Finally, combining Eq.~\ref{eqn:PCuppererror}, Eq.~\ref{eqn:PCcompareerror}, and Eq.~\ref{eqn:PCboundtrue}, we find that:

\begin{align}
\begin{split}
| P^*(C | s, M) & - P(C | s, M)| 
\\ & < \max\{|\overline{P}(C | s, M) - P(C | s, M)|,
\\ & \;\;\;\;\;\;\;\;\;\;\;\;\; | P(C | s, M) - \underline{P}(C | s, M)| \} 
\\ & = \max\{\overline{P}(C | s, M) - P(C | s, M), 
\\ & \;\;\;\;\;\;\;\;\;\;\;\;\; P(C | s, M) - \underline{P}(C | s, M)\} \\
& = \overline{P}(C | s, M) - P(C | s, M) \\
& = \frac{\prod_\theta{|N_\theta(i)|!}}{\prod_{a, \theta}{C(a, \theta)!}}\left[\prod_{a, \theta}{\left(\hat{p}_{a, \theta}+\epsilon_{\hat{p}}\right)^{C(a, \theta)}} - \prod_{a, \theta}{\hat{p}_{a, \theta}^{C(a, \theta)}}\right] 
\\ & = \epsilon_{P(C)} 
\end{split}
\label{eqn:cor1proof}
\end{align}

which establishes Eq.~\ref{eqn:errorPC}. $\square$

\section{Proof for Theorem 2}

Through the following, we establish in Theorem~\ref{thm:regretbound} a bound on the regret of the cumulative discounted rewards that the subject agent $i$ incurs from following a $k$ horizon Many Agent \ipomdplite{} optimal policy $\pi_i$, where the regret is caused by using the \emph{estimated} likelihoods of other agents' configurations $P(C | s, M)$ based on modeling only a few neighbors and extrapolating to the total population (as compared to their \emph{true} likelihoods $P^*(C | s, M)$). \vspace{-0.5em}
\paragraph{Strategy} 

To establish this regret bound, we extend the theoretical analysis of Hoang and Low \cite{Hoang13:Interactive}, following much of the same theoretical framework and arguments that they established (reproduced and adapted here for completeness).  In particular, these extensions below in Lemmas \ref{lem:errorR}-\ref{lem:errorQ}, Proposition~\ref{prop:deltabound}, and Theorem~\ref{thm:regretbound} involve (1) considering more than one other agent in the environment (Hoang and Low only established theoretical results for environments with only two agents), and (2) accounting for the approximation error in $P(C | s, M)$ given by Corollary~\ref{cor:errorPC}.


\begin{lem}[Approximation Error Bound on One Step Rewards]
Let $b_i$ be an arbitrary belief and $a_i$ be an arbitrary action for the subject agent $i$.  Then the expected reward $R_i(b_i, a_i) = E[R_i(b_i, a_i, C)]$ with respect to the \emph{estimated} probabilities of other agents' actions in configurations $P(C | s, M)$, when compared to expected reward $R_i^*(b_i, a_i) = E[R_i(b_i, a_i, C)]$ with respect to the \emph{true} probabilities of other agents' actions $P^*(C | s, M)$, will have bounded error:
\begin{align}
\left|R_i^*(b_i, a_i) - R_i(b_i, a_i)\right| \le \epsilon_{P(C)}|\mathcal{C}|R_{max}
\label{eqn:errorR} 
\end{align}

\noindent which is linear in the bounded approximation error of the configuration probabilities $\epsilon_{P(C)}$, where $\mathcal{C}$ is the set of all possible configurations and $R_{max}$ is the largest reward value for all possible $s, a, C$.
\label{lem:errorR}
\end{lem}

\paragraph{Proof}  This result extends Hoang and Low's Lemma 2 \cite{Hoang13:Interactive}.  Expanding the expected reward values, we have:

\begin{align}
\begin{split}
|R_i^*(b_i, a_i) & - R_i(b_i, a_i)| 
\\ & \le \sum_{s, M}{b_i(s,M) \sum_{C}{R_i(s, a_i, C)}}
\\ & \times |P^*(C | s, M) - P(C | s, M)| \\
& \le \sum_{s, M}{b_i(s, M)\sum_{C}{\epsilon_{P(C)}R_i(s, a_i, C)}} \text{\;\;\;(by Cor.~\ref{cor:errorPC})} \\
& \le \sum_{s, M}{b_i(s, M)\sum_{C}{\epsilon_{P(C)} R_{max}}}  \\
& = \sum_{s, M}{b_i(s, M){\epsilon_{P(C)}|\mathcal{C}| R_{max}}}  \\
& = \epsilon_{P(C)}|\mathcal{C}|R_{max}
\end{split}
\label{eqn:proofErrorR} 
\end{align}

$\square$.


\begin{lem}[Approximation Error Bound on $P_i(C, o_i | b_i, a_i)$]
Let $b_i$ be an arbitrary belief and $a_i$ be an arbitrary action for the subject agent $i$. Then the \emph{estimated} belief-state observation model $P_i(C, o_i | b_i, a_i)$, i.e., the combined likelihood that the other agents perform configuration $C$ and subject agent $i$ observes observation $o_i$, has approximation error compared to its \emph{true} distribution $P_i^*(C, o_i | b_i, a_i)$ at most the bound on the approximation error of $P(C | s, M)$:
\begin{align}
|P_i^*(C, o_i | b_i, a_i) & - P_i(C, o_i | b_i, a_i)| \le \epsilon_{P(C)}
\label{eqn:errorPCO} 
\end{align}
\label{lem:errorPCO}
\end{lem}

\paragraph{Proof}  This result extends Hoang and Low's Lemma 3 \cite{Hoang13:Interactive}.  Expanding the probability functions, we have:

\begin{align}
\begin{split}
|P_i^*(C, o_i | b_i, a_i) & - P_i(C, o_i | b_i, a_i)| 
\\ & \le \sum_{s'}{O_i(s', a_i, C, o_i)\sum_{s,M}{T_i(s, a_i, C, s') b_i(s,M)}}
\\ & \times |P^*(C | s, M) - P(C | s, M)|  \\
& \le \sum_{s'}{O_i(s', a_i, C, o_i)\sum_{s,M}{T_i(s, a_i, C, s')b_i(s,M)}}
\\ & \times \epsilon_{P(C)} \text{\;\;\;(by Cor.~\ref{cor:errorPC})}  \\
& \le \epsilon_{P(C)} \sum_{s, M}{b_i(s, M)\sum_{s'}{O_i(s', a, C, o)T_i(s, a, C, s')}}  \\
& \le \epsilon_{P(C)} \sum_{s, M}{b_i(s, M)\sum_{s'}{T_i(s, a_i, C, s')}}  \\
& = \epsilon_{P(C)} \sum_{s, M}{b_i(s, M)}  \\
& = \epsilon_{P(C)}
\end{split}
\label{eqn:proofErrorPCO} 
\end{align}

$\square$.


\newpage
\begin{lem}[Approximation Error Bound on Belief Updates]
Let $b_i$ be an arbitrary belief, $a_i$ be an arbitrary action and $o_i$ be an arbitrary observation for the subject agent $i$, and let $C$ be an arbitrary configuration of the other agents' actions.  Let $B_i(b_i, a_i, C, o_i)$ represent the belief-update function for subject agent $i$, using it's estimated likelihoods of configurations and producing distribution $P(s' | o_i, a_i, C, b_i)$, and let $F_i(b_i, a_i, C, o_i)$ be its unnormalized form:

\begin{align}
\begin{split}
F_i(b_i, a_i, & C, o_i)(s')
\\ & = O_i(s', a_i, C, o_i) \sum_{s, M}{T_i(s, a_i, C, s') P(C | s, M) b_i(s, M)} 
\\ & = B_i(b_i, a_i, C, o_i) P_i(C, o_i | b_i, a_i)
\end{split}
\label{eqn:unnormbelief} 
\end{align}

\noindent Then the norm-1 distances between $F_i(b_i, a_i, C, o_i)$ and its true values $F_i^*(b_i, a_i, C, o_i)$ calculated with the \emph{true} probabilities of other agents' actions $P^*(C | s, M)$ is also bounded by $\epsilon_{P(C)}$, the approximation error of $P(C | s, M)$:

\begin{align}
\left\|F_i^*(b_i, a_i, C, o_i) - F_i(b_i, a_i, C, o_i)\right\|_1 \le \epsilon_{P(C)}
\label{eqn:errorF}
\end{align}
\label{lem:errorF}
\end{lem}

\paragraph{Proof}  This result extends Hoang and Low's Lemma 5 \cite{Hoang13:Interactive}.  Expanding the norm, we have:

\begin{align}
\begin{split}
||F_i^*(b_i, a_i, C, o_i) & - F_i(b_i, a_i, C, o_i)||_1 
\\ & \le \sum_{s'}{O_i(s', a_i, C, o_i) \sum_{s, M}{T_i(s, a_i, C, s')}}
\\ & \times b_i(s, M) |P^*(C | s, M) - P(C | s, M)|
\\ & \le \sum_{s'}{O_i(s', a_i, C, o_i) \sum_{s, M}{T_i(s, a_i, C, s')}} 
\\ & \times \epsilon_{P(C)} b_i(s, M) \text{\;\;\;(by Cor.~\ref{cor:errorPC})} 
\\ & = \sum_{s, M}{b_i(s, M) \sum_{s'}{O_i(s', a_i, C, o_i) T_i(s, a_i, C, s')}}
\\ & \times \epsilon_{P(C)} 
\\ & \le \sum_{s, M}{b_i(s, M) \sum_{s'}{T_i(s, a_i, C, s')}} 
\\ & \times \epsilon_{P(C)} 
\\ & = \epsilon_{P(C)} \sum_{s, M}{b_i(s, M)} 
\\ & = \epsilon_{P(C)} 
\end{split}
\label{eqn:proofErrorF} 
\end{align}

$\square$.


\vspace{10em} 
\begin{lem}
Let $b_i$ be an arbitrary belief, $a_i$ be an arbitrary action and $o_i$ be an arbitrary observation for the subject agent $i$, and let $C$ be an arbitrary configuration of the other agents' actions.  Let $B_i(b_i, a_i, C, o_i)$ be the belief update function as defined in Lemma 3, calculated using the subject agent $i$'s estimated $P(C | s, M)$, and let $B_i^*(b_i, a_i, C, o_i)$ be similar but calculated using the \emph{true} $P^{*}(C | s, M)$.  Also, let $P_i(C, o_i | b_i, a_i)$ and $P_i^*(C, o_i | b_i, a_i)$ be as given in Lemma 2.  Then:
\begin{align}
\begin{split}
P_i(C, o_i | b_i, a_i) & \left\|B_i^*(b_i, a_i, C, o_i) - B_i(b_i, a_i, C, o_i)\right\|_1 
\\ & \le 2\epsilon_{P(C)}
\end{split}
\label{eqn:error4}
\end{align}

\label{lem:4}
\end{lem}

\paragraph{Proof}  This result extends Hoang and Low's Proposition 1 \cite{Hoang13:Interactive}.  Following Hoang and Low's notation, let $B_i^*$, $B_i$, $F_i^*$, $F_i$, $P_i^*$, and $P_i$ be shorthand for $B_i^*(b_i, a_i, C, o_i)$, $B_i(b_i, a_i, C, o_i)$, $F_i^*(b_i, a_i, C, o_i)$, $F_i(b_i, a_i, C, o_i)$, $P_i^*(C, o_i | b_i, a_i)$, and $P_i(C, o_i | b_i, a_i)$ respectively.  Then, expanding the norm, we have:

\begin{align}
\begin{split}
P_i^* ||B_i^* & - B_i||_1 
\\ & = P_i^* \sum_{s'}{\left|\frac{F_i^*(s')}{P_i^*} - \frac{F_i(s')}{P_i} \right|} \text{\;\;\;\;\; (by Eq.~\ref{eqn:unnormbelief})}
\\ & = \frac{1}{P_i} \sum_{s'}{\left|F_i^*(s') P_i - F_i(s') P_i^* \right|}
\\ & = \frac{1}{P_i} \sum_{s'}{\left| P_i \left(F_i^*(s') - F_i(s')\right) + F_i(s') \left(P_i - P_i^*\right)\right|} 
\\ & \le \frac{1}{P_i} \sum_{s'}{\left[  P_i \left|F_i^*(s') - F_i(s')\right| + F_i(s') \left|P_i^* - P_i \right|\right] } 
\\ & \text{\;\;\;\;\;\;(by Triangle Ineq.)}\\
& \le \frac{1}{P_i} \sum_{s'}{\left[ P_i \left|F_i^*(s') - F_i(s')\right| + F_i(s') \epsilon_{P(C)}\right]} 
\\ & \text{\;\;\;\;\;\;(by Lemma 2)}
\\ & = \sum_{s'}{\left|F_i^*(s') - F_i(s')\right|} + \frac{\epsilon_{P(C)}}{P_i} \sum_{s'}{F_i(s')} 
\\ & = \sum_{s'}{\left|F_i^*(s') - F_i(s')\right|} + \frac{\epsilon_{P(C)}}{P_i} \sum_{s'}{P_i B_i(s')} 
\\ & \text{\;\;\;\;\;\; (by defintion of $F_i$)} \\
& = \sum_{s'}{\left|F_i^*(s') - F_i(s')\right|} + \epsilon_{P(C)} \\
& \le \epsilon_{P(C)} + \epsilon_{P(C)} \text{\;\;\; (by Lemma 3)} \\
& = 2 \epsilon_{P(C)}
\end{split}
\label{eqn:proofError4}
\end{align}

$\square$.

\newpage
\begin{lem}
Let $b_i$ be an arbitrary belief, $a_i$ be an arbitrary action and $o_i$ be an arbitrary observation for the subject agent $i$, and let $C$ be an arbitrary configuration of the other agents' actions.  Let $B_i(b_i, a_i, C, o_i)$ and $B^*(b_i, a_i, C, o_i)$ be as given in Lemma 4, and let $P_i(C, o_i | b_i, a_i)$ and $P_i^*(C, o_i | b_i, a_i)$ be as given in Lemma 2.

Given $P(C | s, M)$, we have:

\begin{align}
\begin{split}
|V_i^*(B_i^* & (b_i, a_i, C, o_i)) P_i^*(C,  o_i | b_i, a_i) 
\\ & \;\;\;\;\;\;- V_i^*(B_i(b_i, a_i, C, o_i)) P_i(C,  o_i | b_i, a_i)|
\\ & \le 3 \epsilon_{P(C)} \frac{R_{max}}{1 - \gamma}
\end{split}
\label{eqn:errorV}
\end{align}

\label{lem:errorV}
\end{lem}

\paragraph{Proof}  This result extends Hoang and Low's Proposition 2 \cite{Hoang13:Interactive}.  As in the proof for Lemma 4, for notational convenience, let $B_i^*$, $B_i$, $P_i^*$, and $P_i$ be shorthand for $B_i^*(b_i, a_i, C, o_i)$, $B_i(b_i, a_i, C, o_i)$, $P_i^*(C, o_i | b_i, a_i)$, and $P_i(C, o_i | b_i, a_i)$ respectively.  Then, expanding the left side of Eq.~\ref{eqn:errorV}, we have:

\begin{align}
\begin{split}
|V_i^*(B_i^*)P_i^* & - V_i^*(B_i)P_i| 
\\ & \le \left|V_i^*(B_i^*)P_i^* - V_i^*(B_i)P_i^*\right| 
\\ & + \left|V_i^*(B_i)P_i^* - V_i^*(B_i)P_i\right| 
\\ & = P_i^* \left|V_i^*(B_i^*) - V_i^*(B_i)\right| + V_i^*(B_i) \left|P_i^* - P_i \right| 
\\ & \le P_i^* \frac{R_{max}}{1 - \gamma} \left\|B_i^* - B_i \right\|_1 + V_i^*(B_i) \left|P_i^* - P_i \right| 
\\ & \text{\;\;\;\;\;(by Hoang and Low (\citeyear{Hoang13:Interactive})'s Lemma 4)}\\
& \le 2 \epsilon_{P(C)} \frac{R_{max}}{1 - \gamma} + V_i^*(B_i) \left|P_i^* - P_i \right| 
\\ & \text{\;\;\;\;\;(by Lemma 4)}
\\ & \le 2 \epsilon_{P(C)} \frac{R_{max}}{1 - \gamma} + V_i^*(B_i) \epsilon_{P(C)}
\\ & \text{\;\;\;\;\;(by Lemma 2)}
\\ & \le 2 \epsilon_{P(C)} \frac{R_{max}}{1 - \gamma} + \frac{R_{max}}{1-\gamma} \epsilon_{P(C)} 
\\ & \text{\;\;\;\;\;(since $V_i^* \le \frac{R_{max}}{1 - \gamma}$)}\\
& = 3 \epsilon_{P(C)} \frac{R_{max}}{1 - \gamma}
\end{split}
\label{eqn:proofErrorV}
\end{align}

$\square$.


\vspace{20em}
\begin{defn}(Maximum Error in Value Function).
Let $V_{i, k}(b_i)$ be the optimal value function calculated using the subject agent $i$'s \emph{estimated} $P(C | s, M)$ likelihoods over other agents' configurations after $k$ backups of the Bellman operator, and let $V_{i.k}^*$ be the similar optimal value function calculated instead using the \emph{true} likelihoods $P^*(C | s, M)$.  Then we define the maximal difference between $V_{i,k}^*$ and $V_{i,k}$, caused by approximating the distribution of configuration likelihoods as:
\begin{align}
\delta_n \triangleq \max_{b} \left|V_{i,n}^*(b) - V_{i,n}(b)\right|
\label{eqn:maxErrorV}
\end{align}
\label{defn:maxErrorV}
\end{defn}

Note that this definition matches that given by Hoang and Low \cite{Hoang13:Interactive} for the I-POMDP-Lite, except their difference is due to approximating the other agents' decision making as a \nestedmdp, whereas our's accounts for extrapolating the models of only a few agents to the collective behavior of the whole system.

\begin{lem}
Let $b_i$ be an arbitrary belief and $a_i$ be an arbitrary action for the subject agent $i$.  Then:

\begin{align}
\begin{split}
|Q_{i,k}^*(b_i, a_i) & - Q_{i, k}(b_i, a_i)| 
\\ & \le \gamma \delta_{k-1} + \epsilon_{P(C)} |\mathcal{C}| R_{max} \left[1 + 3 \gamma \frac{|\Omega_i|}{1 - \gamma} \right]
\end{split}
\label{eqn:errorQ}
\end{align}

\noindent where $Q_{i, n}(b_i, a_i)$ is the the expected $Q$ value over all possible configurations $\mathcal{C}$ using the subject agent's \emph{estimated} $P(C | s, M)$ multinomial distribution after $k$ backups of the Bellman operation; $Q_{i, k}^*$ is the similar Q value calculated instead using the \emph{true} likelihoods $P^*(C | s, M)$, and $\delta_{k-1}$ is as defined in Def.~\ref{defn:maxErrorV}.

\label{lem:errorQ}
\end{lem}

\paragraph{Proof}  This result extends Hoang and Low's Proposition 3 \cite{Hoang13:Interactive}.  As in the proofs for Lemmas ~\ref{lem:errorF} - ~\ref{lem:errorV}, let us use $B_i$ and $B_i^*$ as shorthand notation.  Analogous to Hoang and Low's Proposition 3, we define:
\begin{align}
L_{i, k}(b_i, a_i) = R(b_i, a_i) + \gamma \sum_{C}{\sum_{o_i}{V_{i, k-1}^*(B_i)P_i(C, o_i | b_i, a_i)}}
\label{eqn:lk}
\end{align}

\noindent Looking first at $\left|L_{i, k}(b_i, a_i) - Q_{i,k}(b_i, a_i)\right|$, we establish:

\begin{align}
\begin{split}
 |L_{i, k}(b_i, a_i) & - Q_{i,k}(b_i, a_i) |  
 \\ & \le \gamma \sum_{C}{\sum_{o_i}{P_i(C, o_i | b_i, a_i) \left|V_{i, k-1}^*(B_i) - V_{i, k}(B_i) \right|}} 
 \\ & \le \gamma \sum_{C}{\sum_{o_i}{P_i(C, o_i | b_i, a_i) \delta_{k-1}}} 
 \\ & \le \gamma \delta_{k-1} \sum_{C}{\sum_{o_i}{P_i(C, o_i | b_i, a_i)}} 
 \\ & = \gamma \delta_{k-1}
\end{split}
\label{eqn:LminusQ}
\end{align}

\newpage
\noindent Looking next at $\left|Q_{i,k}^*(b_i, a_i) - L_{i, k}(b_i, a_i)\right|$, we establish:

\begin{align}
\begin{split}
 |Q_{i,k}^*(b_i, a_i) & - L_{i, k}(b_i, a_i)|
 \\ & \le \left| R_i^*(b_i, a_i) - R_i(b_i, a_i) \right| 
 \\ & + \gamma \sum_{C}{\sum_{o_i}{|V_{i, k-1}^*(B_i^*)P_i^*(C, o_i | b_i, a_i)}}
 \\ & \;\;\;\;\;\;\;\;\;\;\;\;\;\;\;\;\;\;\; - V^*_{i, k}(B_i)P_i(C, o_i | b_i, a_i)| 
 \\ & \le \left| R_i^*(b_i, a_i) - R_i(b_i, a_i) \right| 
 \\ & + \gamma \sum_{C}{\sum_{o_i}{3 \epsilon_{P(C)} \frac{R_{max}}{1-\gamma}}} \text{\;\;\;\;(by Lemma 5)}
 \\ & \le \epsilon_{P(C)} |\mathcal{C}| R_{max} + \gamma \sum_{C}{\sum_{o_i}{3 \epsilon_{P(C)} \frac{R_{max}}{1-\gamma}}} \\ & \text{\;\;\;\;\;(by Lemma 1)}
 \\ & = \epsilon_{P(C)} |\mathcal{C}| R_{max} + 3 \gamma \epsilon_{P(C)} |\mathcal{C}| |\Omega_i|  \frac{R_{max}}{1-\gamma} 
 \\ & = \epsilon_{P(C)} |\mathcal{C}| R_{max} \left[1 + 3\gamma \frac{|\Omega_i|}{1-\gamma} \right]
\end{split}
\label{eqn:QminusL}
\end{align}

\noindent Finally, by the Triangle Inequality and combining Eq.~\ref{eqn:LminusQ} and Eq.~\ref{eqn:QminusL}, we have:

\begin{align}
\begin{split}
|Q_{i,k}^*(b_i, a_i) & - Q_{i, k}(b_i, a_i)| 
\\ & \le |Q_{i,k}^*(b_i, a_i) - L_{i, k}(b_i, a_i)|
\\ & + | L_{i, k}(b_i, a_i) - Q_{i,k}(b_i, a_i)| 
\\ & \le \gamma \delta_{k-1} + \epsilon_{P(C)} |\mathcal{C}| R_{max} \left[1 + 3 \gamma \frac{|\Omega_i|}{1 - \gamma} \right]
\end{split}
\label{eqn:proofErrorQ}
\end{align}

$\square$.


\begin{prop}
The maximum difference $\delta_{k}$ between the optimal value function $V_{i, k}^*$ calculated using the \emph{true} distribution over other agents' configurations $P^*(C | s, M)$ and the approximated optimal value function $V_{i,k}$ from the \emph{estimated} distribution $P(C | s, M)$ is bounded by:

\begin{align}
\delta_{k} \le \epsilon_{P(C)} |\mathcal{C}| R_{max} \left[ \gamma^{k-1} + \frac{1}{1-\gamma}\left(1 + 3\gamma \frac{|\Omega_i|}{1 - \gamma} \right) \right]
\label{eqn:deltabound}
\end{align}

which is bounded \emph{linearly} by $\epsilon_{P(C)}$ and \emph{decreases} as $k$ (i.e., the number of Bellman backups) increases, approaching an asymptote of $\epsilon_{P(C)} |\mathcal{C}| R_{max} \frac{1}{1-\gamma}\left(1 + 3\gamma \frac{|\Omega_i|}{1 - \gamma} \right)$

\label{prop:deltabound}
\end{prop}

\paragraph{Proof} This result extends Hoang and Low's Proposition 4 \cite{Hoang13:Interactive}.  Based on Hoang and Low's Lemma 1 \cite{Hoang13:Interactive}, we have that:

\begin{align}
\delta_{k} \le \max_{b_i} \max_{a_i} \left| Q_{i, k}^*(b_i, a_i) - Q_{i, k}(b_i, a_i) \right|
\label{eqn:deltaboundQ}
\end{align}

\noindent Thus, we find that:

\begin{align}
\begin{split}
\delta_{k} & \le \max_{b_i} \max_{a_i} \left| Q_{i, k}^*(b_i, a_i) - Q_{i, k}(b_i, a_i) \right| 
\\ & \le \gamma \delta_{k-1} + \epsilon_{P(C)} |\mathcal{C}| R_{max} \left[1 + 3 \gamma \frac{|\Omega_i|}{1 - \gamma} \right] \text{(by Lemma 6)} 
\\ & \le \gamma^{k-1} \delta_{1} + \epsilon_{P(C)} |\mathcal{C}| \frac{R_{max}}{1 - \gamma} \left[1 + 3 \gamma \frac{|\Omega_i|}{1 - \gamma} \right] 
\\ & \text{\;\;\;\;\;(by unrolling the Bellman recurrence)} 
\\ & \le \gamma^{k-1} \epsilon_{P(C)} |\mathcal{C}| R_{max} + \epsilon_{P(C)} |\mathcal{C}| \frac{R_{max}}{1 - \gamma} \left[1 + 3 \gamma \frac{|\Omega_i|}{1 - \gamma} \right] 
\\ & \text{\;\;\;\;\;(by Lemma 1)} 
\\ & = \epsilon_{P(C)} |\mathcal{C}| R_{max} \left[ \gamma^{k-1} + \frac{1}{1-\gamma}\left(1 + 3\gamma \frac{|\Omega_i|}{1 - \gamma} \right) \right]
\end{split}
\label{eqn:proofdeltabound}
\end{align}

\noindent This establishes that $\delta_{k}$ is linearly bounded by $\epsilon_{P(C)}$ (Eq. ~\ref{eqn:deltabound}).  Furthermore, since $\gamma \in [0, 1)$, we have that $\gamma^{k-1}$ approaches 0 as $k$ increases.  Therefore, as $k$ increases, the $\gamma^{k-1}$ term disappears and $\delta_{k}$ approaches the tigher bound of $\epsilon_{P(C)} |\mathcal{C}| R_{max} \frac{1}{1-\gamma}\left(1 + 3\gamma \frac{|\Omega_i|}{1 - \gamma} \right)$.
$\square$


\begin{defn}(Actual Return Earned by Subject Agent).
Let $J_{i, k}(b_i)$ represent the \emph{actual} expected cumulative discounted rewards earned by the subject agent $i$ over a horizon of $k$ if it follows its optimal policy $\pi_i$ calculated using the \emph{estimated} likelihoods of configurations of other agents' actions $P(C | s, M)$, even though they chose their actions in the environment according to the \emph{true} likelihood $P^*(C | s, M)$:

\begin{align}
\begin{split}
J_{i, k}(b_i) {} & = \sum_{s, M}{b_i(s, M) \sum_{C}{P^*(C | s, M) R_i(s, \pi_i(b_i), C)}} 
\\ & + \gamma \sum_{C}{b_i(s, M) \sum_{o_i}{P_i^*(C, o_i | b_i, \pi_i(b_i))}}
\\ & \;\;\;\;\;\;\;\;\;\;\;\; \times J_{i, k-1}(B_i(b_i, \pi_i(b_i), C, o_i))
\label{eqn:actualreturn}
\end{split}
\end{align}

\noindent where $B_i$ represents the belief update function for subject agent $i$ and $P_i^*(C, o_i | b_i, \pi_i(b_i))$ represents the true likelihood that the other agents choose configuration $C$ and the subject agent observes $o_i$ after taking action $\pi_i(b_i)$ with belief current $b_i$.
\label{defn:actualreturn}
\end{defn}

\newpage
\setcounter{equation}{7}
\begin{thm}(Regret bound).
Maximum regret that agent $i$ incurs $\left\| V_{i, k}^* - J_{i, k} \right\|_\infty$ from following a $k$-horizon optimal policy $\pi_i$ (obtained by solving the many-agent I-POMDP-Lite) due to the approximate likelihoods of other agents' configurations $P(C | s^t, M^t)$ is bounded from above:
\begin{align}
\begin{split}
\left\| V_{i, k}^* - J_{i, k} \right\|_\infty & \le 2 \epsilon_{P(C)}\cdot |\mathcal{C}| \cdot R_{max} \\ & \times \left[\gamma^{k-1} + \frac{1}{1-\gamma} \left(1 + 3\gamma\frac{|\Omega_i|}{1 - \gamma}\right) \right]
\end{split}
\label{eqn:regretboundx2}
\end{align}

\noindent which is linear in the error $\epsilon_{P(C)}$ in the subject agent's estimation in configuration likelihoods caused by only modeling some neighbors.
\label{thm:regretbound2}
\end{thm}

\setcounter{equation}{41}
\paragraph{Proof} This result extends Hoang and Low's Theorem 4 \cite{Hoang13:Interactive}.  For notational convenience, define the maximal difference between the subject agent $i$'s computed value function (using its estimated likelihoods $P(C | s, M)$) and its actual expected return:
\begin{align}
\phi_k \triangleq \max_b |V_{i, k}(b) - J_{i, k}(b)|
\label{eqn:x}
\end{align}

\noindent Finally, as in prior lemmas, we use $B_i^*$ and $B_i$ as shorthand notation for $B_i^*(b_i, \pi_i(b_i), C, o_i)$ and $B_i(b_i, \pi_i(b_i), C, o_i)$, respectively.

\noindent To begin, we establish several intermediate results that will be useful to derive Eq.~\ref{eqn:regretbound}.  First, we find that:

\begin{align}
\begin{split}
P_i^*(C, o_i & | b_i, \pi_i(b_i)) \left| J_{i, k-1}(B_i^*) - V_{i, k-1}(B_i) \right| 
\\ & \le P_i^*(C, o_i | b_i, \pi_i(b_i)) \left| J_{i, k-1}(B_i^*) - V_{i, k-1}(B_i^*) \right| 
\\ & + P_i^*(C, o_i | b_i, \pi_i(b_i)) \left| V_{i, k-1}(B_i^*) - V_{i, k-1}(B_i) \right| 
\\ & \le P_i^*(C, o_i | b_i, \pi_i(b_i)) \left| J_{i, k-1}(B_i^*) - V_{i, k-1}(B_i^*) \right| 
\\ & + P_i^*(C, o_i | b_i, \pi_i(b_i)) \frac{R_{max}}{1-\gamma} \left\| B_i^* - B_i \right\|_1 
\\ & \; \text{(by Hoang and Low (\citeyear{Hoang13:Interactive})'s Lemma 4)}
\\ & \le P_i^*(C, o_i | b_i, \pi_i(b_i)) \left| J_{i, k-1}(B_i^*) - V_{i, k-1}(B_i^*) \right| 
\\ & + 2 \epsilon_{P(C)} \frac{R_{max}}{1-\gamma} \text{\;\;(by Lemma 4)} 
\\ & \le P_i^*(C, o_i | b_i, \pi_i(b_i)) \phi_{k-1} + 2 \epsilon_{P(C)} \frac{R_{max}}{1-\gamma}
\end{split}
\label{eqn:thm2part1}
\end{align}

\noindent Next, we find that:
\begin{align}
\begin{split}
|P_i^*(C, o_i & | b_i, \pi_i(b_i)) J_{i, k-1}(B_i^*) - P_i(C, o_i | b_i, \pi_i(b_i)) V_{i, k-1}(B_i)| 
\\ & \le P_i^*(C, o_i | b_i, \pi_i(b_i)) \left| J_{i, k-1}(B_i^*) - V_{i, k-1}(B_i) \right| 
\\ & + V_{i, k-1}(B_i) \left| P_i^*(C, o_i | b_i, \pi_i(b_i)) - P_i(C, o_i | b_i, \pi_i(b_i))  \right| 
\\ & \le P_i^*(C, o_i | b_i, \pi_i(b_i)) \left| J_{i, k-1}(B_i^*) - V_{i, k-1}(B_i) \right| 
\\ & + V_{i, k-1}(B_i) \epsilon_{P(C)} \text{\;\; (by Lemma 2)} 
\\ & \le P_i^*(C, o_i | b_i, \pi_i(b_i)) \left| J_{i, k-1}(B_i^*) - V_{i, k-1}(B_i) \right| 
\\ & + \epsilon_{P(C)} \frac{R_{max}}{1 - \gamma} \text{\;\;\;\;\;\;\;\;\; (since $V_{i, k-1} \le \frac{R_{max}}{1 - \gamma}$)} 
\\ & \le P_i^*(C, o_i | b_i, \pi_i(b_i)) \phi_{k-1} + 2 \epsilon_{P(C)} \frac{R_{max}}{1-\gamma} 
\\ & + \epsilon_{P(C)} \frac{R_{max}}{1 - \gamma} \text{\;\;\;\;\;\;\;\;\;\;(by Eq.~\ref{eqn:thm2part1})}
\\ & = P_i^*(C, o_i | b_i, \pi_i(b_i)) \phi_{k-1} + 3 \epsilon_{P(C)} \frac{R_{max}}{1-\gamma} 
\end{split}
\label{eqn:thm2part2}
\end{align}

\noindent Following, we find that:
\begin{align}
\begin{split}
|V_{i, k}(b_i) & - J_{i, k}(b_i)| 
\\ & \le \sum_{s, M}{b_i(s, M) \sum_{C}{R_i(s, \pi_i(b_i), C)}}
\\ & \;\;\;\;\;\;\;\;\;\;\;\;\;\;\;\;\;\;\;\;\;\;\;\;\;\;\;\;\;\; \times \left| P^*(C | s, M) - P(C | s, M) \right| \\
& + \gamma \sum_{C}{\sum_{o_i}{| P_i^*(C, o_i | b_i, \pi_i(b_i)) J_{i, k-1}(B_i^*)}}
\\ & \;\;\;\;\;\;\;\;\;\;\;\;\;\;\;\;\;\;\;\; - P_i(C, o_i | b_i, \pi_i(b_i)) V_{i, k-1}(B_i)| \\
& \le \epsilon_{P(C)} |\mathcal{C}| R_{max} \text{\;\;\;\;\;\;\;\;\;\;\;\; (by Lemma 1)} \\
& + \gamma \sum_{C}{\sum_{o_i}{| P_i^*(C, o_i | b_i, \pi_i(b_i)) J_{i, k-1}(B_i^*)}}
\\ & \;\;\;\;\;\;\;\;\;\;\;\;\;\;\;\;\;\;\;\; - P_i(C, o_i | b_i, \pi_i(b_i)) V_{i, k-1}(B_i)|
\\ & \le \epsilon_{P(C)} |\mathcal{C}| R_{max} 
\\ & + \gamma \sum_{C}{\sum_{o_i}{\left[ P_i^*(C, o_i | b_i, \pi_i(b_i)) \phi_{k-1} + 3 \epsilon_{P(C)} \frac{R_{max}}{1-\gamma} \right] }} 
\\ & \text{\;\;\;\;\;(by Eq.~\ref{eqn:thm2part2})}
\\ & = \epsilon_{P(C)} |\mathcal{C}| R_{max} 
\\ & + \gamma \sum_{C}{\sum_{o_i}{P_i^*(C, o_i | b_i, \pi_i(b_i)) \phi_{k-1} }} 
\\ & + 3 \gamma |\mathcal{C}| |\Omega_i| \epsilon_{P(C)} \frac{R_{max}}{1-\gamma} 
\\ & \le \epsilon_{P(C)} |\mathcal{C}| R_{max} + 3 \gamma |\mathcal{C}| |\Omega_i| \epsilon_{P(C)} \frac{R_{max}}{1-\gamma}  + \gamma \phi_{k-1}
\end{split}
\label{eqn:thm2part3}
\end{align}

\newpage
\noindent Rewriting this as a recurrence relation based on the definition above of $\phi_k$, we have:

\begin{align}
\begin{split}
\phi_k {} & \le \epsilon_{P(C)} |\mathcal{C}| R_{max} + 3 \gamma |\mathcal{C}| |\Omega_i| \epsilon_{P(C)} \frac{R_{max}}{1-\gamma}  + \gamma \phi_{k-1} 
\\ & = \gamma \phi_{k-1} + \epsilon_{P(C)} |\mathcal{C}| R_{max} \left[1 +  3 \gamma \frac{|\Omega_i|}{1-\gamma} \right] 
\\ & \le \gamma^{k-1} \phi_{1} + \epsilon_{P(C)} |\mathcal{C}| R_{max} \frac{1}{1 - \gamma} \left( 1 +  3 \gamma \frac{|\Omega_i|}{1-\gamma} \right) 
\\ & \text{\;\;\;\;\;(by unrolling the recurrence)} 
\\ & \le \gamma^{k-1} \epsilon_{P(C)} |\mathcal{C}| R_{max} 
\\ & + \epsilon_{P(C)} |\mathcal{C}| R_{max} \frac{1}{1 - \gamma} \left( 1 +  3 \gamma \frac{|\Omega_i|}{1-\gamma} \right) 
\\ & \text{\;\;\;\;\;(by Lemma 1)} \\
& = \epsilon_{P(C)} |\mathcal{C}| R_{max} \left[ \gamma^{k-1}  + \frac{1}{1 - \gamma} \left( 1 +  3 \gamma \frac{|\Omega_i|}{1-\gamma} \right) \right] 
\end{split}
\label{eqn:thm2part4}
\end{align}

\noindent Hence, we must have that:
\begin{align}
\begin{split}    
|V_{i, k}(b_i) & - J_{i, k}(b_i)| 
\\ & \le \phi_k 
\\ & \le \epsilon_{P(C)} |\mathcal{C}| R_{max} \left[ \gamma^{k-1}  + \frac{1}{1 - \gamma} \left( 1 +  3 \gamma \frac{|\Omega_i|}{1-\gamma} \right) \right] 
\\ & \text{\;\;\;\;\;(by Eq.~\ref{eqn:thm2part4})}
\end{split}
\label{eqn:thm2part5}
\end{align}

\noindent Finally, we can establish our main result in Eq.~\ref{eqn:regretbound}.  We find that for arbitrary belief $b_i$:
\begin{align}
\begin{split}
|V_{i, k}^*(b_i) & - J_{i, k}(b_i)|
\\ & \le \left|V_{i, k}^*(b_i) - V_{i, k}(b_i) \right| + \left|V_{i, k}(b_i) - J_{i, k}(b_i) \right| 
\\ & \le \epsilon_{P(C)} |\mathcal{C}| R_{max} \left[ \gamma^{k-1}  + \frac{1}{1 - \gamma} \left( 1 +  3 \gamma \frac{|\Omega_i|}{1-\gamma} \right) \right] 
\\ & \text{\;\;\;\;\;(by Prop.~\ref{prop:deltabound})} 
\\ & + \left|V_{i, k}(b_i) - J_{i, k}(b_i) \right| 
\\ & \le \epsilon_{P(C)} |\mathcal{C}| R_{max} \left[ \gamma^{k-1}  + \frac{1}{1 - \gamma} \left( 1 +  3 \gamma \frac{|\Omega_i|}{1-\gamma} \right) \right] 
\\ & + \epsilon_{P(C)} |\mathcal{C}| R_{max} \left[ \gamma^{k-1}  + \frac{1}{1 - \gamma} \left( 1 +  3 \gamma \frac{|\Omega_i|}{1-\gamma} \right) \right] 
\\ & \text{\;\;\;\;\;(by Eq.~\ref{eqn:thm2part5})} 
\\ & = 2 \epsilon_{P(C)} |\mathcal{C}| R_{max} \left[ \gamma^{k-1}  + \frac{1}{1 - \gamma} \left( 1 +  3 \gamma \frac{|\Omega_i|}{1-\gamma} \right) \right] 
\end{split}
\label{eqn:thm2part6}
\end{align}

\noindent Since Eq.~\ref{eqn:thm2part6} holds for arbitrary $b_i$, we conclude that:
\begin{align}
\left\| V_{i, k}^* - J_{i, k} \right\|_\infty \le 2 \epsilon_{P(C)} |\mathcal{C}| R_{max} \left[\gamma^{k-1} + \frac{1}{1-\gamma} \left(1 + 3\gamma\frac{|\Omega_i|}{1 - \gamma}\right) \right]
\label{eqn:regretbound2}
\end{align}
which is a restatement of Eq.~\ref{eqn:regretbound}.
$\square$.

\section{Additional Empirical Results}
Here we present the second and third performance measures: the average number of fire locations put out and the average amount of suppressant used by agents.  They are included here as there was not sufficient room in the manuscript.

\begin{figure}[!ht]
\centering
\includegraphics[width=3.25in]{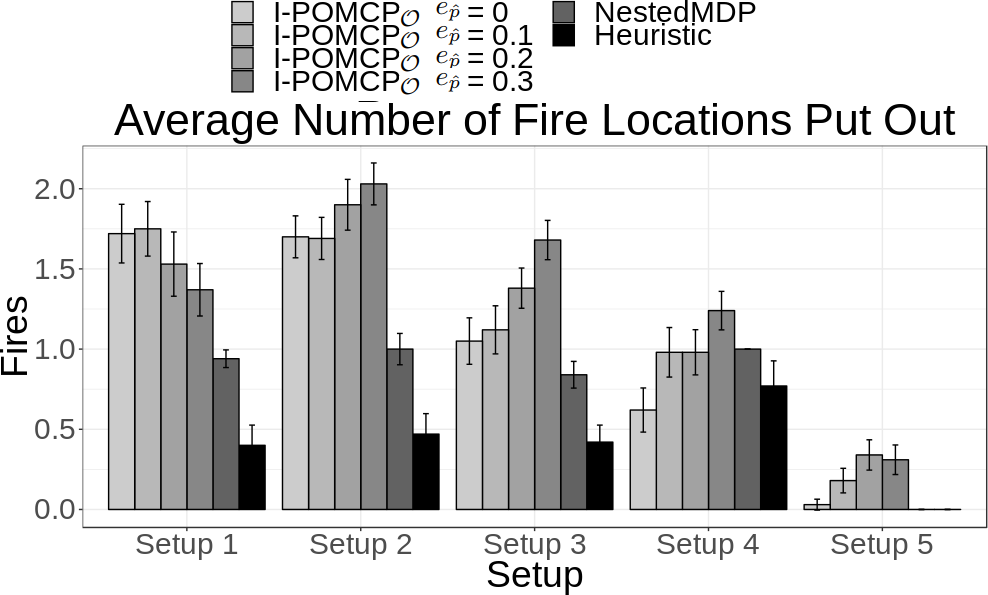}
\caption{Average number of fire locations put out per run across 100 runs. Error bars represent the 95\% confidence interval. I-POMCP planned for 5 seconds in Setups 1-3 and 5, and for 20 seconds in Setup 4.}
\label{fig:firesout}
\vspace{-0.2in}
\end{figure}

\begin{figure}[!ht]
\centering
\includegraphics[width=3.25in]{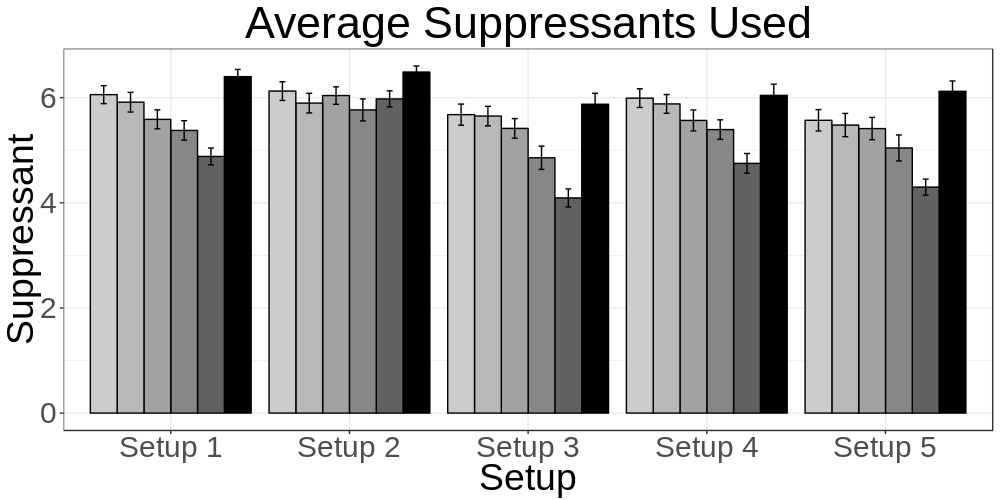}
\caption{Average units of suppressant used per agent averaged across 100 runs. Error bars represent the 95\% confidence interval. I-POMCP planned for 5 seconds in Setups 1-3 and 5, and for 20 seconds in Setup 4.}
\label{fig:suppressants}
\vspace{-0.2in}
\end{figure}

\end{document}